# MnSnTeO$_6$: a Chiral Antiferromagnet Prepared by a Two-Step Topotactic Transformation

*Elena Zvereva$^{1*}$, Kirill Bukhteev$^1$, Maria Evstigneeva$^2$, Evgenia Komleva$^3$, GrigoryRaganyan$^1$, Konstantin Zakharov$^1$, Yevgeny Ovchenkov$^1$, Alexander Kurbakov$^{4,5}$, Mariia Kuchugura$^{4,5}$, Anatoliy Senyshyn$^6$, SergeyStreltsov$^{3,7}$, Alexander Vasiliev$^{1,8,9}$, Vladimir Nalbandyan$^2$*

$^1$ Faculty of Physics, Moscow State University, Moscow, 119991, Russia
$^2$ Faculty of Chemistry, Southern Federal University, Rostov-on-Don 344090, Russia
$^3$ Institute of Metal Physics, Ekaterinburg, 620990, Russia
$^4$ NRC «Kurchatov Institute» - PNPI, Gatchina 188300, Russia
$^5$ Faculty of Physics, St. Petersburg University, St. Petersburg 198504, Russia
$^6$ Heinz Maier-Leibnitz Zentrum, Technische Universität München, Garching 85748, Germany
$^7$ Ural Federal University, Ekaterinburg 620002, Russia
$^8$ National Research South Ural State University, Chelyabinsk 454080, Russia
$^9$ National University of Science and Technology "MISiS", Moscow 119049, Russia

**ABSTRACT**

MnSnTeO$_6$, a new chiral antiferromagnet, was prepared both by topotactic transformation of the metastable rosiaite-type polymorph and by direct synthesis from coprecipitated hydroxides. Its structure, static and dynamic magnetic properties were studied comprehensively both experimentally (through X-ray and neutron powder diffraction, magnetization, specific heat, dielectric permittivity and ESR techniques) and theoretically (by means of *ab initio* density functional theory (DFT) calculations within the spin-polarized generalized gradient approximation). MnSnTeO$_6$ is isostructural with MnSb$_2$O$_6$ (space group *P*321) and does not show any structural transition between 3 and 300 K. The magnetic susceptibility and specific heat exhibit an antiferromagnetic ordering at $T_N \approx 9.8$ K, which is confirmed by low-temperature neutron data. At the same time, the thermodynamic parameters demonstrate an additional anomaly on the temperature dependences of magnetic susceptibility $\chi(T)$, specific heat $C_p(T)$ and dielectric permittivity $\varepsilon(T)$ at $T^* \approx 4.9$ K, which is characterized by significant temperature hysteresis. Clear enhancement of the dielectric permittivity at $T^*$ is most likely to reflect the coupling of dielectric and magnetic subsystems leading to development of electric polarization. It was established that the ground state of MnSnTeO$_6$ is stabilized by seven exchange parameters, and neutron diffraction revealed incommensurate magnetic structure with propagation vector ***k***= (0, 0, 0.183) analogous to that of MnSb$_2$O$_6$. *Ab initio* DFT calculations demonstrate that the strongest exchange coupling occurs between planes along diagonals. All exchange parameters are antiferromagnetic and reveal moderate frustration.

## 1. INTRODUCTION

For the long time, the family of chiral magnets was limited to intermetallic compounds based on MnSi and FeGe, crystallizing in space group $P2_13$[1-5] and the iron langasite Ba$_3$NbFe$_3$Si$_2$O$_{14}$, crystallizing in space group *P*312.[6,7] Spiral (helical) magnetic order develops in these materials for $T < T_c$, which amounts to $\approx 29$ K for MnSi, $\approx 280$ K for FeGe and 27 K for Ba$_3$NbFe$_3$Si$_2$O$_{14}$. Hierarchy of exchange interactions includes isotropic symmetric exchange interactions as a major counterpart, the antisymmetric Dzyaloshinskii-Moriya interactions (DMI) on the intermediate scale and higher-order spin–orbit coupling on the weakest scale. The DMI is triggered by the noncentrosymmetric crystal structure and stabilizes helical magnetic order in these compounds. In magnetic field, for MnSi, two another phases occur. Thus, magnetic phase diagram comprises the helical, conical, and skyrmion lattice states.[8] MnSi was the first alloy where unique topological skyrmion phase (a hexagonal lattice of spin vortices) has been discovered by means of neutron scattering, which was followed by real space maps of these particle-like states using Lorentz transmission electron microscopy (TEM)[1,3]. Two more examples of intermetallic compounds with skyrmion, which is worth to mention, are Cr$_{1/3}$NbS$_2$[9] and the family of Co-Zn-Mn alloys.[10]

Two different types of particle-like states (solitons) were predicted in skyrmion phases: vortex-like spin textures—skyrmion tubes (SkTs) and the chiral bobber (ChB).[7] Quite recently, the direct experimental observation of ChBs in thin films of B20-type FeGe has been reported using a quantitative off-axis electron holography.[8] It was also shown that ChBs are able to coexist with SkTs over a wide range of parameters. Obviously, it is potentially interesting for practical applications in creation of nonvolatile memory devices since two types of these quasiparticles can be used for encoding binary data bits as a sequence of skyrmions and bobbers.

First *insulating* material where skyrmions have been discovered was multiferroic compound $Cu_2OSeO_3$.[11-14] $Cu_2OSeO_3$ crystallize into the same $P2_13$ space group symmetry as MnSi alloys. The magnetic structure is ferromagnetic with two sublattices corresponding to two different positions of $Cu^{2+}$ ions in the structure: Cu sites in a square pyramidal and trigonal bipyramidal oxygen environments. The investigation of $Cu_2OSeO_3$ indicated that the sample develops a spontaneous ferroelectric polarization along the [111] direction in response to an applied magnetic field in the same direction. This means some degree of magnetoelectric coupling.[11,12] The lack of conductivity in this compound indicates rather different mechanism of stabilization of skyrmionic phase compared to the conductive materials.

Later, other insulators among chiral magnets have been revealed. Examples are $CuB_2O_4$ with the tetragonal space group $I$-$42d$[15,16] and $MnSb_2O_6$ with trigonal $P321$ symmetry[17-19]. $CuB_2O_4$ demonstrated successive magnetic ordering through canted antiferromagnetic commensurate phase 10 K < $T$ < 21 K towards an incommensurate helical phase at low temperatures.[15] Besides, $CuB_2O_4$ revealed giant optical magnetoelectric effect[16] which involves very low-energy magnetic excitations.

Multidomain structure has been reported for $MnSb_2O_6$.[18,19] According to Johnson et al.,[18] the $MnSb_2O_6$ magnetic structure is based on corotating cycloids rather than helices unlike all other known cases. $MnSb_2O_6$ is predicted to be weakly polar and multiferroic with unique mechanism of switching of the electric polarization leading to a switch between single- and two-domain configurations. Detailed thermodynamic and resonance studies revealed a crucial role of the planar anisotropy for the spin structure of $MnSb_2O_6$.[19] It is reflected by competing antiferromagnetic phases arising in moderate magnetic field, which most probably favor helical spin structure over the cycloidal ground state.

Unusual physical properties of $MnSb_2O_6$[18-20] stimulate interest in preparation and investigation of its structural analogues. However, $MnSb_2O_6$ seems to be an extremely rare magnetic representative of its structure type, at least among oxides. In the $MSb_2O_6$ series, all $M^{2+}$ cations larger than $Mn^{2+}$ (M = Cd, Ca, Sr, Pb, and Ba) give rise to the rosiaite structure type whereas all stable compounds with $M^{2+}$ cations smaller than $Mn^{2+}$ (M = Mg, Co, Ni, Cu, and Zn) belong to the trirutile structure type,[21,22] although metastable rosiaite-type polymorphs were also prepared.[22] All known $M^{2+}As_2O_6$ (M =Ni, Co, Mn, Pd, Ca, Pb) are also rosiaites.[23-26] Both trirutile and rosiaite types are centrosymmetric and, thus, less interesting than $MnSb_2O_6$. Substitution of X = Nb or Ta for Sb, despite close similarity of ionic radii, also results in different $M^{2+}X_2O_6$ structure types.[21,27,28] To our knowledge, among different $M^{2+}X_2O_6$ compounds, only $CoU_2O_6$ and $NiU_2O_6$ crystallize in the $MnSb_2O_6$ structure type, although with different arrangement of $M^{2+}$ and $U^{5+}$.[29-31]

To find non-radioactive structural analogues of $MnSb_2O_6$, heterovalent substitutions should be attempted. Of these, the one most similar to $MnSb_2O_6$ should be its isoelectronic counterpart, $MnSnTeO_6$. However, its direct solid-state synthesis from the corresponding binary oxides would be very difficult because activation of $SnO_2$ requires very high temperatures at which Te(6+) compounds might be unstable and volatile. In this work, we applied two different approaches to preparation of $MnSnTeO_6$ and we report for the first time on synthesis and detailed study of magnetic properties of the new member of chiral magnets, $MnSnTeO_6$, which belongs to the same structure type with trigonal space group $P321$ as $MnSb_2O_6$.

When this work was in progress, we discovered a half-century-old report[32] mentioning $MnSnTeO_6$ isostructural with $MnSb_2O_6$. However, the preparation method, crystal structure details and physical properties were not reported there; of the two reported hexagonal lattice parameters, *a* agrees reasonably with our data but *c* is 5 % larger than ours. The reason for this discrepancy is not clear and no further publication on this material appeared.

In present paper we have comprehensively studied new chiral ($P321$) $MnSnTeO_6$ both experimentally and theoretically. For the sake of proper comparison and to outline the difference in magnetic properties with previously studied rosiaite ($P\bar{3}1m$) polymorph[35] we have also performed additional measurements for the latter phase. This paper is organized as follows:

Section 2 contains experimental details.

Section 3 describes the main results and discussion and is divided into 10 subsections: 3.1. Samples identification; 3.2. Structural analysis; 3.3. Magnetic susceptibility; 3.4. Specific heat, 3.5. Dielectric permittivity; 3.6. Magnetization isotherms; 3.7. ESR spectroscopy; 3.8. Magnetic phase diagram; 3.9. Magnetic structure; 3.10. Density functional calculations: electronic structure and exchange interaction parameters.

Section 4 contains concluding remarks.

## 2. EXPERIMENTAL SECTION
### 2.1. Sample preparation and elemental analysis

Two different approaches to preparation of $MnSnTeO_6$ were applied. First, we designed a two-step preparation route analogous to our preceding work with $MnSb_2O_6$[33] as illustrated in Fig. 1. It includes a low-temperature topotactic ion-exchange preparation of a rosiaite-type phase, $MnSb_2O_6$ or $MnSnTeO_6$, from $Na_2Sb_2O_6$ or $Na_2SnTeO_6$, respectively, followed by its topotactic transformation to the $P321$ phase upon heating. This transformation may proceed in an intact crystal because the two structures are very similar. Both are based on the hexagonal eutaxy of oxygen anions with cations filling half octahedral voids and the only difference is the allocation of one ninth of the cations. In the $P321$ structure of $MnSb_2O_6$, alternating cationic layers are $Mn_3Sb$ and $Sb_5$ whereas in the metastable rosiaite-type $MnSb_2O_6$ ($P\bar{3}1m$), they are $Mn_3$ and $Sb_6$ or merely Mn and $Sb_2$. Thus, the transformation only requires displacement of 1/9 of the cations from one layer to another. This results in vanishing of both mirror plane and inversion center and appearance of superlattice reflections indicating multiplication of the hexagonal lattice constant $a$ by a factor of $\sqrt{3}$ whereas the constant $c$ remains almost unchanged.

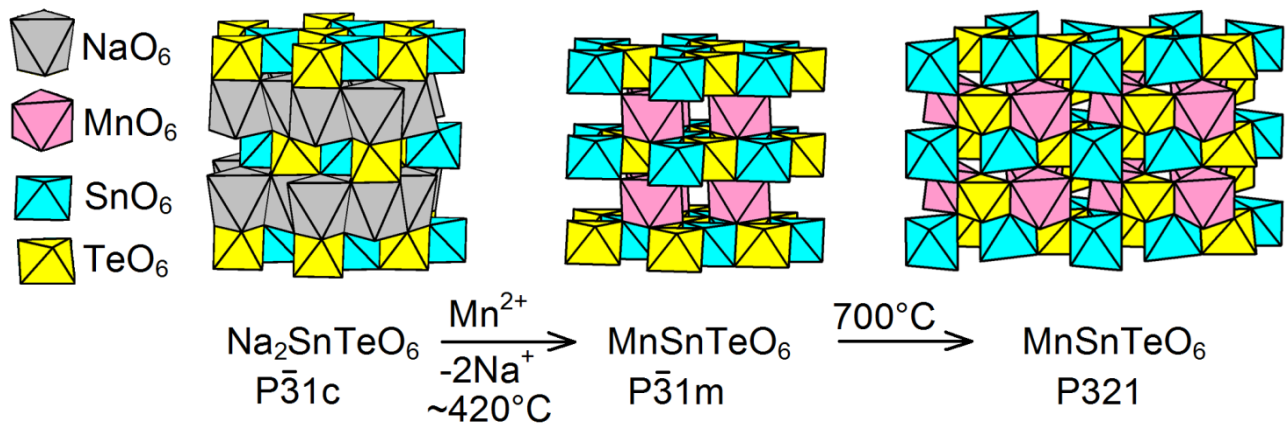

**Figure 1.** Polyhedral presentation of crystal structures and their transformations, left to right: $Na_2SnTeO_6$;[34] metastable rosiaite-type $MnSnTeO_6$ (the model with disordered stacking of completely ordered $SnTeO_6^{2-}$ layers);[35] stable $MnSnTeO_6$ (this work).

The ion-exchange preparation of the metastable rosiaite-type $MnSnTeO_6$ in molten salt mixtures was reported earlier.[35] After washing out the salts and subsequent drying, the product was heated to somewhat higher temperatures for conversion to the stable chiral polymorph. Our alternative route to chiral $MnSnTeO_6$ included co-precipitation of Mn and Sn hydroxides from aqueous solution followed by their solid-state reaction with $TeO_3$ or $TeO_2$ in air. Details of each preparation may be found in the Supporting Information (pdf).

For elemental analysis, an electron microprobe (INCA ENERGY 450/XT) was used with an X-Act ADD detector based on an electron microscope VEGA II LMU (Tescan) operated at the accelerating voltage of 20 kV. The molar ratios Mn/Te, Sn/Te and Na/Te were averaged on six measurements at various points of the sample.

## 2.2. Diffraction studies

For X-ray powder diffraction (XRPD) measurements, we used an ARL X'TRA diffractometer equipped with a solid-state Si(Li) detector eliminating almost completely all undesired wavelengths and leaving essentially pure CuKα doublet. Lattice parameters were refined with CELREF 3 (J. Laugier & B. Bochu) after angular corrections with an internal standard, corundum NIST SRM 676. For structural refinements by the Rietveld method, the GSAS + EXPGUI suite[36,37] was used.

Neutron powder diffraction (NPD) measurements were carried out on the diffractometer SPODI[38] in Heinz Maier-Leibnitz Zentrum (MLZ), Garching, Germany in the temperature range from 3 K to 10 K ($T$ = 3, 5, 7, 8, 10 K). At $T$ = 15 and 300 K, i.e. in the paramagnetic state of the sample, the measurements were performed with lower statistics compared to low-temperature measurements, when magnetic order was of primary interest. Reflection (551) from a germanium crystal monochromatorat 155° take off angle was used, which corresponds to the wavelength of the incident neutrons $\lambda$ = 1.548 Å and the high-resolution mode of the diffractometer.

The sample was placed in a vanadium container with a diameter of 6 mm. Neutron diffraction patterns were recorded in the angular interval $2\theta$ = 1 - 150 deg. with a step of 0.05 deg. A closed-cycle refrigerator was used. The model of the crystal structure obtained at 15 K in the paramagnetic state was the basic model of the refinement of the experimental neutron diffraction patterns measured at lower temperatures to obtain the parameters of the magnetic structure.

The full-profile refinement of the experimental neutron diffraction patterns was carried out by the Rietveld method using the Fullprof suite.[39] Graphic construction of the models of crystal and magnetic structures was carried out using the VESTA,[40] and Jmolcodes.[41] The magnetic reflections were indexed with the help of the *k*-Search program (in the Fullprof suite). The analysis of the irreducible representations of the space groups and the basis functions corresponding to the magnetic moments of the atoms as axial vectors, was carried out for the modelling of the magnetic structure using the BasIreps program (in the Fullprof suite).

Detailed structural data (experimental conditions, discrepancy factors, atomic coordinates, displacement parameters, individual bond lengths, etc.) may be found in the Supporting Information (pdf and cif), and the main text only includes lattice parameters, average bond lengths, experimental and calculated patterns and graphical presentations of the crystal and magnetic structure.

## 2.3. Magnetic and dielectric measurements

Magnetic measurements were performed by means of a Quantum Design MPMS XL-7 magnetometer. The temperature dependences of the magnetic susceptibility were measured at the magnetic field $B$ = 0.1 T in the temperature range 1.8–300 K. The isothermal magnetization curves were obtained in static magnetic fields $B \leq 7$ T and at $T \leq 20$ K after cooling the sample in zero magnetic field.

Magnetic measurements in pulsed magnetic fields were carried out in a temperature range 2.5 – 12 K using a 30 T system consisting of a multi-turn solenoid and a capacitor bank providing a rise time of about 8 ms. The isotherms $M(B)$ were measured with a conventional pick-up magnetometer by numerical integration of the signal which induces the moment of the sample in the compensated coil during a magnetic field pulse.

The specific heat measurements were carried out by a relaxation method using a Quantum Design PPMS-9 system. The plate-shaped sample of ~0.2 mm thickness and ~8 mg, mass was obtained by cold pressing the powder. Data were collected at zero magnetic field and under applied fields up to 9 T in the temperature range 2 – 30 K.

The dielectric constant was measured on a Quantum Design MPMS XL-7 system using a custom-made insert and Andeen-Hagerling 2700A capacitive bridge. The measurements were performed at various frequencies in the range 1-20 kHz at temperatures from 2 to 300 K in zero and applied magnetic fields up to 7 T. Resolution of Andeen-Hagerling 2700A is 2.4 - 16 aF depending on frequency. In addition, the device can measure the loss tangent down to $1.5 \times 10^{-8}$, conductivity up to $3 \times 10^{-7}$ nS, or resistance up to $1.7 \times 10^{6}$ GOhm. The duration of one measurement was from 30 ms to 0.4 s. Operating voltage was 15 V. The dielectric permittivity sample was a plane-parallel plate with a diameter of up to 6 mm and a thickness of up to 3 mm. Silver paste "LietSilber" was applied to it on both sides, resulting in a capacitor. The sample was mounted in a measuring insert for the MPMS XL-7 system, which was used to control temperature and apply an external magnetic field.

In order to get the full set of data for comparison and to outline the difference in physical properties of two polymorphs we have also measured the specific heat and dielectric permittivity for metastable rosiaite-type MnSnTeO$_6$ sample since these data were not studied or reported earlier.[35] The measurements conditions were used the same as described above.

Electron spin resonance (ESR) studies were carried out using an X-band ESR spectrometer CMS 8400 (ADANI) ($f \approx 9.4$ GHz, $B \leq 0.7$ T) equipped with a low-temperature mount, operating in the range $T = 6$–300 K. The main spectroscopic parameters are: microwave power 1 mW, magnetic field modulation frequency 100 kHz and amplitude 1 mT and spectral resolution 0.15 mT/pt. The effective $g$-factors of studied sample have been calculated with respect to an external reference for the resonance field. The BDPA ($a,g$ - bisdiphenyline-$b$-phenylallyl) $g_{ref} = 2.00359$ was used as a reference material.

## 3. RESULTS AND DISCUSSION
### 3.1. Sample identification

The ion-exchange preparation of small portions of the metastable rosiaite-type MnSnTeO$_6$ was reported earlier.[35] Preparation of a larger portion, necessary for the neutron diffraction, required longer duration or somewhat higher temperature to melt the salt mixture taken in a five-fold excess. As a result, we could not avoid partial dissolution-precipitation of the product, and it contained, besides the main rosiate-type phase, two additional phases: SnO$_2$ and the stable chiral ($P$321) polymorph. SnO$_2$ most probably resulted from incongruent dissolution of Na$_2$SnTeO$_6$ in the melt because tin dioxide is the most refractory of all components.

Fig. 2a shows XRPD pattern of the purest rosiaite-type sample obtained by ion exchange at 430 °C,[35] and Fig. 2b, the result of calcination of a similar sample (but with SnO$_2$ impurity) for 1 h at 650 °C. This demonstrates that transformation to the stable phase in absence of molten salts is incomplete and requires somewhat higher temperatures, thus confirming the dissolution-precipitation mechanism. At 700 °C, however, the solid-state transformation is essentially complete, and Fig. 2c shows the XRPD pattern of the largest $P$321 sample (Sample 1, about 4.3 g), containing unavoidable admixture of SnO$_2$ and used for neutron diffraction and other experiments in this work. The minor impurity phase (SnO$_2$) was included into both X-ray and neutron refinement and its content was found to be in good agreement: 4.67 and 5.06 wt. %, respectively.

The EDX analysis of Sample 1 yielded the following molar ratios: Mn/Te = 1.04(3), Sn/Te = 1.05(4) and Na/Te = 0.07(4). The first two values are in reasonable agreement with the nominal composition, especially taking into account the content of SnO$_2$ as a separate phase. One of the analyzed points showed much larger Sn content (obviously due to a grain of SnO$_2$) and was excluded from the averaging analyses. Considerable content of sodium seems, however, strange. It cannot be explained by incomplete Mn substitution because the found Mn content is slightly elevated rather than deficient. Therefore, we conclude that sodium in Sample 1 exists most probably as adsorbed or occluded residual salts.

Samples 2 (Fig. 2d) and 3 were prepared by direct solid-state synthesis. They represent the same phase as Sample 1. However, their unit cell volumes are 0.4-0.7 % smaller (Table 1) and, again, small admixture of SnO$_2$ is detected (much smaller in Sample 3). These results indicated that the phase under study has somewhat variable composition. Although investigation of its homogeneity range was out of the scope of the present work, Sample 1 required more attention because it was used for all studies described below (Samples 2 and 3 appeared much later), and was shown to be essentially stoichiometric (see Section 3.2).

Table 1 demonstrates analogy between lattice parameters of MnSnTeO$_6$ and MnSb$_2$O$_6$ in both structure types. The average octahedral ionic radius[42] for the Sn$^{4+}$+ Te$^{6+}$ couple is 0.025 Å bigger than that for Sb$^{5+}$. Thus, it is quite natural that the unit cell volume of chiral MnSnTeO$_6$ is ~1% bigger than that for MnSb$_2$O$_6$. With the rosiaite-type forms, however, the difference is negligible. This may be attributed to admixture of Na in rosiaite-type MnSb$_2$O$_6$.[33] In both cases, the lattice parameter changes on SnTe substitution for Sb$_2$ are anisotropic but, unexpectedly, opposite in sign: in the $P$321 type, they are -0.2% and +1.4% for $a$ and $c$, respectively, whereas the corresponding changes are +0.5% and -1.1% in the rosiaite type.

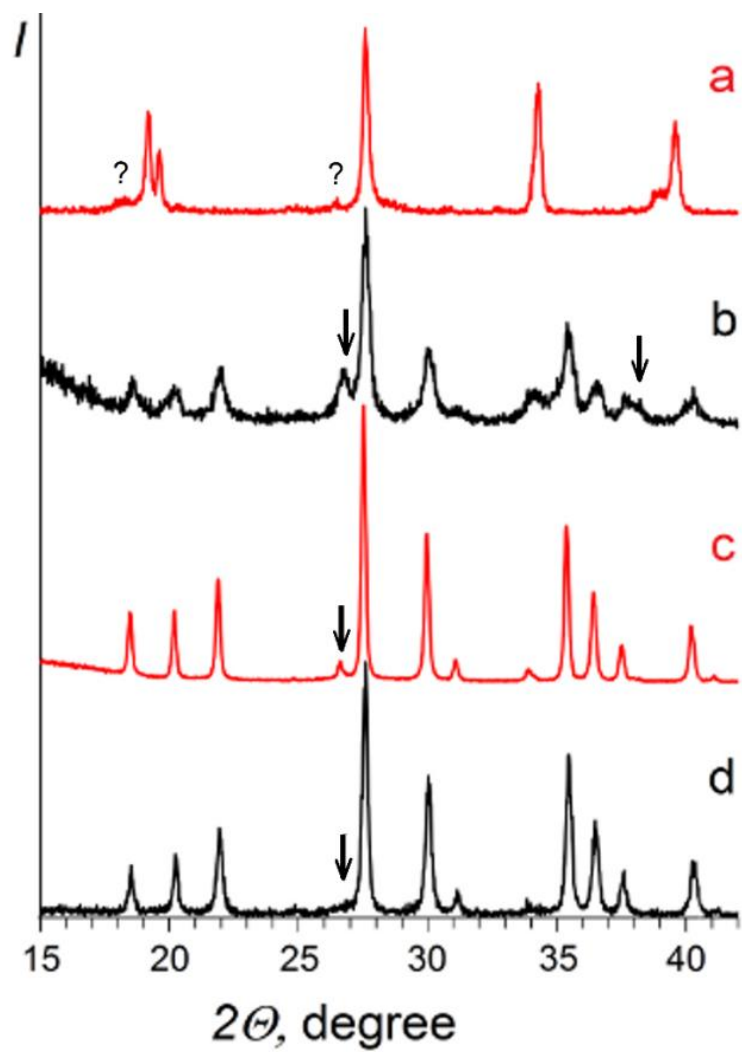

**Figure 2**. Room-temperature XRPD patterns of various MnSnTeO$_6$ preparations (see Section 3.1 and Supporting Information for details).
(a) Rosiaite-type;[35] (b) Similar sample with SnO$_2$ impurity calcined for 1 h at 650 °C; (c) Sample 1: the largest ion-exchange preparation calcined for 1 h at 700 °C, $P321$ form + SnO$_2$; (d) Sample 2: solid-state preparation, 4 h at 750 °C. Arrows, reflections from SnO$_2$; question marks, unknown reflections.

### 3.2. Structural analysis

Atomic scattering factors of tin and tellurium are very similar not only for X-rays ($Z_{Sn}$ =50, $Z_{Te}$ =52), but also for neutrons (coherent scattering lengths are $b_{Sn}$= 0.62·10$^{-12}$ cm and $b_{Te}$=0.58·10$^{-12}$ cm). Therefore, it is difficult to differentiate directly between tin and tellurium from diffraction data. Nevertheless, we managed to solve this problem unambiguously using considerable difference (0.13 Å) in their octahedral ionic radii.[42]
In our initial structural model for the X-ray study within space group $P321$, Sn was placed on sites 1a and 2d, Te on 3f and Mn, on 3e. Refinement of this model showed, however, that bond distances to oxygen on both "Sn" sites are considerably shorter than those on the "Te" site in contrast with ionic sizes. Thus, these two components were interchanged and the structure was successfully refined. Quite similar results were obtained from neutron diffraction at 15 K. At this temperature, the compound is still in a paramagnetic state and the neutron scattering from the sample is purely nuclear.

Experimental and calculated XRPD profiles are compared in Fig. 3, refined lattice constants from XRPD and NPD data are listed in Table 1, and more detailed structural data may be found in Tables S1, S2 and S3 of the Supporting Information. Average bond lengths agree well with corresponding sums of ionic radii; bond valence sums are also reasonable (Table 2). Similar to the metastable rosiaite-type form,[35] cationic environment of the three independent oxygen anions is almost planar, sums of the three bond angles being between 354° and 358°.

**Table 1.** Hexagonal lattice parameters of various $MnSnTeO_6$ samples in comparison with those for $MnSb_2O_6$. For description of samples 1, 2, and 3, see Section 3.1. The value $c = 5.049$ Å[32] is an obvious error. All XRPD data are for room temperature whereas the NPD data are for 15 K.

| | $a$, Å | $c$, Å | $V$, Å$^3$ |
|---|---|---|---|
| $MnSnTeO_6$, rosiaite type, refined on 24 X-ray reflections with internal standard[35] | 5.2313(20) | 4.6109(2) | 109.3 |
| $MnSb_2O_6$, rosiaite type[33] | 5.20649(9) | 4.66276(15) | 109.46 |
| $MnSb_2O_6$, $P321$ [17] | 8.8011(3) | 4.7241(1) | 316.90 |
| $MnSb_2O_6$, $P321$ [32] | 8.802 | 4.719 | 316.6 |
| $MnSnTeO_6$, $P321$[32] | 8.781 | 5.049 | 337.2 |
| $MnSnTeO_6$, $P321$, Sample 1, refined on 44 X-ray reflections with internal standard | 8.7822(8) | 4.7928(1) | 320.13 |
| Same, XRPD Rietveld refinement | 8.78366(10) | 4.79333(9) | 320.27 |
| Same, NPD Rietveld refinement at 15 K | 8.77444(6) | 4.78464(4) | 319.02 |
| $MnSnTeO_6$, $P321$, Sample 2, refined on 34 X-ray reflections with internal standard | 8.7608(18) | 4.7838(1) | 318.0 |
| $MnSnTeO_6$, $P321$, Sample 3, refined on 23 X-ray reflections with internal standard | 8.768(2) | 4.7865(2) | 318.7 |

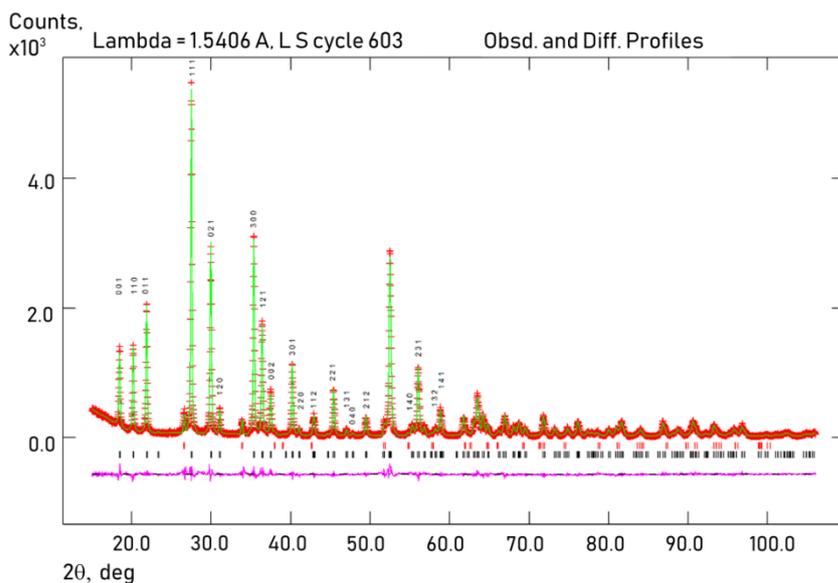

**Figure 3.** Rietveld refinement of XRPD data measured at room temperature. Red crosses are experimental data, green line represents calculated intensity, magenta line is the difference curve, ticks are the position of Bragg reflections for $MnSnTeO_6$ (black) and impurity $SnO_2$ (red). Note that only those reflections where $h^2+hk+k^2 \neq 3n$ (e.g., 011, 021, 120, etc.) cannot be indexed with the smaller rosiaite-type cell and are specific for the larger $P321$ cell.

Variations in lattice parameters mentioned above indicated variable composition. The most realistic model of nonstoichiometry was $Mn_{1-x}Sn_{1-x}Te_{1+x}O_6$ and, thus, Mn occupancy was refined. Due to the well-known correlations between site occupancies and thermal displacement parameters, the 15 K NPD data were preferred for this task. The results shown in Table S4 mean that, within experimental accuracy, Sample 1 is essentially stoichiometric, in agreement with the elemental analysis and bond valence data.

Table 3 compares cationic distribution in the novel triple oxide with those in known double oxides of the same structure type. It is obvious that, irrespective of oxidation state and electronic structure, the smallest cation occupies sites 1a and 2d, the largest one prefers 3e whereas the rest, either small or large, is located on 3f site.

**Table 2**. Average bond lengths and bond valence sums (BVS[43]) in the stable form of $MnSnTeO_6$, space group $P321$, in comparison with the sums of ionic radii.

| Atom | Te1 | Te2 | Sn | Mn | O1 | O2 | O3 |
|---|---|---|---|---|---|---|---|
| Wyckoff site | 1a | 2d | 3f | 3e | 6g | 6g | 6g |
| Average distance to O, Å, XRPD at 298 K | 1.916 | 1.972 | 2.060 | 2.163 | | | |
| Same, NPD at 15 K | 1.934 | 1.899 | 2.058 | 2.197 | | | |
| BVS according to the XRPD at 298 K | 6.02 | 5.17 | 3.95 | 2.20 | 1.98 | 1.91 | 1.97 |
| Same according to the NPD at 15 K | 6.36 | 5.82 | 3.99 | 2.04 | 2.04 | 1.99 | 1.98 |
| Average distance to O, Å, NPD at 10 K | 1.943 | 1.907 | 2.051 | 2.202 | | | |
| Average distance to O, Å, NPD at 3 K | 1.946 | 1.902 | 2.051 | 2.193 | | | |
| Sum of radii,[42] Å | 1.92 | 1.92 | 2.05 | 2.19 | | | |

Polyhedral presentations of the two $MnSnTeO_6$ polymorphs is compared in Fig. 1, whereas Fig. 4 shows cationic distribution in the two consecutive octahedral layers of the $P321$ structure. It is evident now that it is Te that "drops" from one layer to another (0 0 1/2 to 0 0 0) during the transition from rosiaite $P\bar{3}1m$ to chiral $P321$.

**Table 3**. Distribution of cations on four octahedral sites in six isostructural compounds. Shannon's octahedral radii[42] (Å) are given in parentheses.

| Wyckoff site, space group $P321$ | 1a | 2d | 3e | 3f |
|---|---|---|---|---|
| $MnSb_2O_6$[17] | $Sb^{5+}$ (0.74) | $Sb^{5+}$ (0.74) | $Mn^{2+}$ (0.97) | $Sb^{5+}$ (0.74) |
| $CoU_2O_6$[29,30] | $Co^{2+}$ (0.885) | $Co^{2+}$ (0.885) | $U^{5+}$ (0.90) | $U^{5+}$ (0.90) |
| $NiU_2O_6$[29-31] | $Ni^{2+}$ (0.83) | $Ni^{2+}$ (0.83) | $U^{5+}$ (0.90) | $U^{5+}$ (0.90) |
| $MnSnTeO_6$ | $Te^{6+}$ (0.70) | $Te^{6+}$ (0.70) | $Mn^{2+}$ (0.97) | $Sn^{4+}$ (0.83) |
| $Tl_2TeO_6$[44] | $Te^{6+}$ (0.70) | $Te^{6+}$ (0.70) | $Tl^{3+}$ (1.025) | $Tl^{3+}$ (1.025) |
| $Yb_2TeO_6$[45] | $Te^{6+}$ (0.70) | $Te^{6+}$ (0.70) | $Yb^{3+}$ (1.008) | $Yb^{3+}$ (1.008) |

**Table 4**. Mn-Mn distances (Å) around each $Mn^{2+}$ ion in $MnSnTeO_6$ and $MnSb_2O_6$.

| Experiment | along $c$ | in the (001) plane |
|---|---|---|
| $MnSnTeO_6$, XRPD, 298 K | 4.793×2 | 4.814×4, 5.636×2 (17.1 % difference) |
| $MnSnTeO_6$, NPD, 15 K | 4.796×2 | 4.797×4, 5.658×2 (17.9 % difference) |
| $MnSnTeO_6$, NPD, 10 K | 4.785×2 | 4.803×4, 5.645×2 (17.5 % difference) |
| $MnSnTeO_6$, NPD, 3 K | 4.785×2 | 4.804×4, 5.643×2 (17.5 % difference) |
| $MnSb_2O_6$, NPD, 298 K[17] | 4.724×2 | 4.844×4, 5.596×2 (15.5 % difference) |
| $MnSb_2O_6$, NPD, 14 K[17] | 4.721×2 | 4.878×4, 5.519×2 (13.1 % difference) |

A triangular sublattice of magnetic manganese ions, which is special interest for us, is located only in every second layer (the first, third, etc.) with $SnO_6$ - $TeO_6$ layers in between. Fig. 4 might make illusion that each $Mn^{2+}$ ion has six $Mn^{2+}$ neighbors at approximately equal distances within the layer. In fact, however, the $Mn^{2+}$ ions are displaced from the gravity centers of their oxygen octahedra as shown by arrows in Fig. 4, obviously due to repulsion from $Te^{6+}$. As a result, the six Mn-Mn distances within the $Mn_3Te$ layer are essentially different, and the same is observed in the $Mn_3Sb$ layer of $MnSb_2O_6$[17] (Table 4). However, the shortest Mn-Mn distances in both structures are in the perpendicular direction where $Mn^{2+}$ ions form straight chains along the $c$ axis.

Note that the same situation is observed in the next layer, where a similar shift of Sn ions is observed.

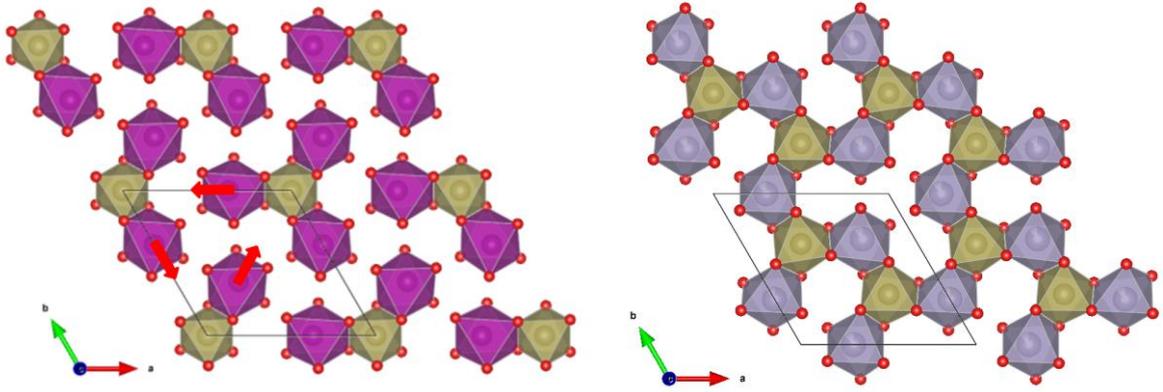

**Figure 4.** Polyhedral view of two consecutive layers in the crystal structure of MnSnTeO$_6$. The first layer (z=0, left) is formed by MnO$_6$ and TeO$_6$ octahedra, magenta and yellow, respectively. Displacement of Mn ions is shown by red arrows. The second layer (z=1/2, right) consists of SnO$_6$ and TeO$_6$ octahedra, violet and yellow, respectively. O$^{2-}$ ions are small red spheres. Rhombus formed by straight lines indicates the crystallographic unit cell.

### 3.3. Magnetic susceptibility

The temperature dependence of the static magnetic susceptibility $\chi = M/B$ of MnSnTeO$_6$ is shown in Fig. 5. There is no visible difference between zero-field cooled (ZFC) and field cooled (FC) modes of $\chi(T)$ signaling absence of either spin-glass or cluster glass effects. At elevated temperature the $\chi(T)$ exhibits a maximum at $T_{max} \approx 10.5$ K, which is likely to indicate approaching a long-range antiferromagnetic ordered state at low temperatures. The Neel temperature as inferred from $d\chi/dT(T)$ is $T_N = 9.8 \pm 0.2$ K. The difference in the values of the $T_{max}$ and $T_N$ is typical for low-dimensional magnets, where the role of short-range order correlations is very high and according to Fisher[43] the maximum at $d(\chi_{\parallel}T)/dT(T)$ should corresponding anomaly in the specific heat. Approximation of the $\chi(T)$ in the temperature range from 200 to 300 K using the modified Curie Weiss law $\chi(T) = \chi_0 + C/(T - \Theta)$ gives the values of the Curie constant $C = 4.23$ emu/mol K and the Weiss temperature $\Theta \approx -46.1$ K.

Remarkably, the Weiss temperature is more than twice larger than the value reported for rosiaite-type ($P\bar{3}1m$) MnSnTeO$_6$, $\Theta = -20.7 \pm 1$ K.[35] This implies not only predominance of the antiferromagnetic correlations, but also increased role of the spin frustration ($f = \Theta/T_N \approx 5$) in the system under study in relevance with a chiral triangular network of Mn$^{2+}$ ions in magnetically active layers of $P321$ MnSnTeO$_6$.

To reduce the number of variable parameters, the value $\chi_0$ has been estimated independently by summation of the Pascal constants[47] for the diamagnetic contributions of ions constituting MnSnTeO$_6$ and the obtained value $\chi_0 = -1.14 \times 10^{-4}$ emu/mol has been fixed. The effective magnetic moment, estimated from the Curie constant as high as $\mu_{eff} = 5.82$ $\mu_B$/f.u., is in satisfactory agreement with the theoretical estimate in accordance with $\mu_{theor} = [ng^2S(S+1)]^{0.5}$ for the high-spin Mn$^{2+}$($S = 5/2$), using $g = 1.986 \pm 0.005$, obtained experimentally from the ESR data (see Section 3.7).

### 3.4. Specific heat

A careful analysis of the magnetic susceptibility in comparison with its derivative $d\chi/dT$ reveals the presence of additional anomaly at $T^* \approx 4.9$ K for new chiral MnSnTeO$_6$ polymorph (left panel of Fig. S1 in the Supporting Information). In contrast, the previous studies on metastable $P\bar{3}1m$ rosiaite-type MnSnTeO$_6$ did not show such a feature (right panel of Fig. S1 in Supporting Information). At the same time, both polymorphs demonstrate obvious difference between the maximum on the $\chi(T)$ and the maximum on its derivative $d\chi/dT(T)$. Such a reduction is typical for antiferromagnets with short-range interactions, i.e. for all low-dimensional magnets as has been shown by Fisher.[48]

The presence of both critical regions at the same temperature range for new $P321$ MnSnTeO$_6$ polymorph is even more evident from the specific heat data. The temperature dependence of the specific heat $C_p(T)$ in a zero magnetic field demonstrates two distinct anomalies, which apparently have the different nature (Fig. 6a). The first one, λ-type anomaly, is observed at $T_N$ = 9.9 K and is typical for the second order transition to 3D antiferromagnetic state. At the same time, there is a broad step-like feature on the $C_p(T)$ which manifests a hysteresis when recording temperature dependence on heating and cooling. Associated critical temperature estimated from the middle of the hysteresis range nicely matches the $T^*$ detected from the $\chi(T)$ data (left panel of Fig. S1 in the Supporting Information). $T$-hysteresis behavior is usually characteristic of the first-order transition but is not limited to. In the magnetic fields, the position of both these anomalies slightly shifts towards lower temperatures (Fig. 7). The $C_p(T)$ behavior of rosiaite-type ($P\bar{3}1m$) MnSnTeO$_6$ is obviously different: it only shows λ-type anomaly at $T_N$ = 8.8 K without temperature hysteresis (Fig. 6b).

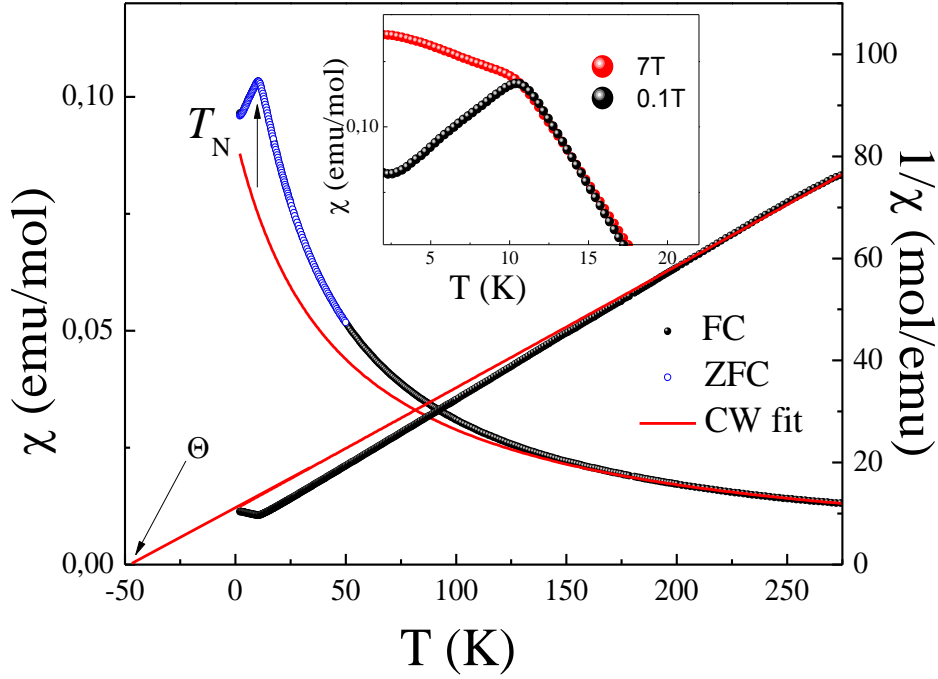

**Figure 5.** Temperature dependence of the magnetic susceptibility of MnSnTeO$_6$ at $B$ = 0.1 T recorded in FC and ZFC cooling modes, as well as the inverse magnetic susceptibility $1/\chi$. The solid red curve is an approximation in accordance with the Curie-Weiss law. On insert: the effect of magnetic field.

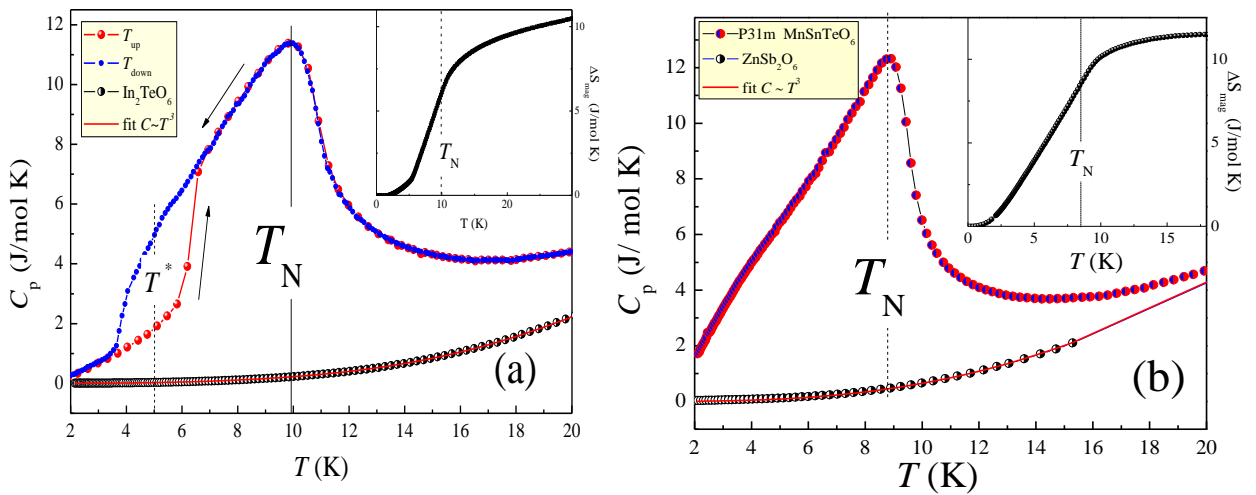

**Figure 6.** (a) The temperature dependence of the specific heat $C_p(T)$ of chiral ($P321$) MnSnTeO$_6$ when recording in the heating mode (red symbols) and in the cooling mode (blue symbols) and its diamagnetic analogue In$_2$TeO$_6$ (black half-filled symbols); (b) The temperature dependence of the specific heat $C_p(T)$ of rosiaite ($P\bar{3}1m$) MnSnTeO$_6$ (red/blue symbols) and its diamagnetic analogue ZnSb$_2$O$_6$ (black half-filled symbols). On the insets: a change in the magnetic entropy.

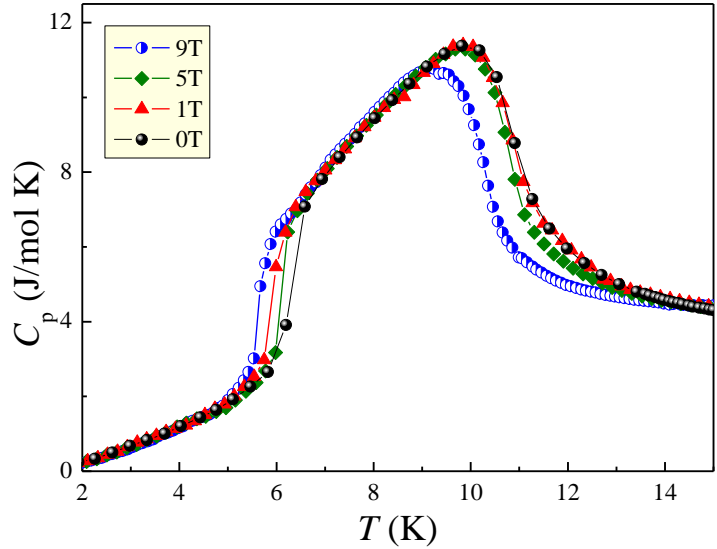

**Figure 7.** The $C_p(T)$ of chiral ($P321$) MnSnTeO$_6$ under the variation of the magnetic field.

In order to evaluate the phonon contribution and estimate the magnetic counterpart to the specific heat and the entropy at the $T_N$ and $B = 0$ T for both MnSnTeO$_6$ polymorphs, the specific heat of the relevant isostructural non-magnetic analogues, In$_2$TeO$_6$ and ZnSb$_2$O$_6$, were measured and subtracted from the data. The jumps in the specific heat for both samples, i.e. $\Delta C_{\mathrm{magn}} \approx 10.9$ J/(mol K) (for $P321$) and $\Delta C_{\mathrm{magn}} \approx 11.9$ J/(mol K) (for $P\bar{3}1m$) do not reach the value expected from the mean-field theory $\Delta C_{\mathrm{theor}} = 5RS(S+1)/[S^2 + (S+1)^2] = 19.6$ J/mol K were $R$ being the gas constant 8.31 J/mol K. As one can see from the insets of Fig. 6, the entropy changes associated with the evolution of magnetic order, i.e. $\Delta S_{\mathrm{magn}} \approx 10.6$ J/(mol K) (for $P321$) and $\Delta S_{\mathrm{magn}} \approx 11.5$ J/(mol K) (for $P\bar{3}1m$), are markedly lower than the total magnetic entropy saturation $\Delta S_{\mathrm{magn}} = R\ln(2S+1)$ expected for system with Mn$^{2+}$ ($S=5/2$). Moreover, for both polymorphs, at $T_N$, less than 65% of the entropy was found to be released. This indicates the presence of appreciable short-range correlations far above $T_N$, which is usually a characteristic feature for materials with low magnetic dimensionality and frustration.[48,49]

### 3.5. Dielectric permittivity

Further peculiarities of the physical properties of chiral ($P321$) MnSnTeO$_6$ were revealed from the dielectric studies. The temperature dependencies of the dielectric constant of MnSnTeO$_6$ at selected frequencies are summarized in upper panel of Fig. 8. Once again we also added the similar data for metastable $P\bar{3}1m$ rosiaite-type MnSnTeO$_6$ for comparison (lower panel of Fig. 8). At all investigated frequencies, there is a rapid increase in $\varepsilon$ near room temperature for both polymorphs. The observed increase in the permittivity at high temperatures can be explained by the appearance of free charge carriers caused by intrinsic defects (for example vacancies and interstitials). The data show no anomaly in $\varepsilon$ at magnetic ordering temperature $T_N$. At the same time, both the real and imaginary parts of the dielectric constant of chiral ($P321$) MnSnTeO$_6$ exhibit a step-like anomaly in the low-temperature region, which is observed at approximately 4.7 K (inset (a) on upper panel of Fig. 8). Remarkably, the position of this anomaly reasonably agree with the second anomaly at $T^* \approx 4.9$ K which was detected from the magnetic susceptibility and specific heat data. Moreover, overall behavior of the dielectric constant is well correlated with the behavior of the specific heat and it also exhibits a pronounced hysteresis in the low-temperature region (inset (a) on upper panel of Fig. 8). Despite the smallness of the absolute magnitude of this effect, its presence is reproduced for all frequencies studied from 1 to 20 kHz. Magnetic nature of the anomaly at $T^* \approx 4.9$ K is confirmed from the effect of magnetic field which suppresses it as shown on inset (b) on upper panel of Fig. 8. In contrast, there are no additional anomalies on the permittivity of metastable $P\bar{3}1m$ rosiaite-type MnSnTeO$_6$ (lower panel of Fig. 8).

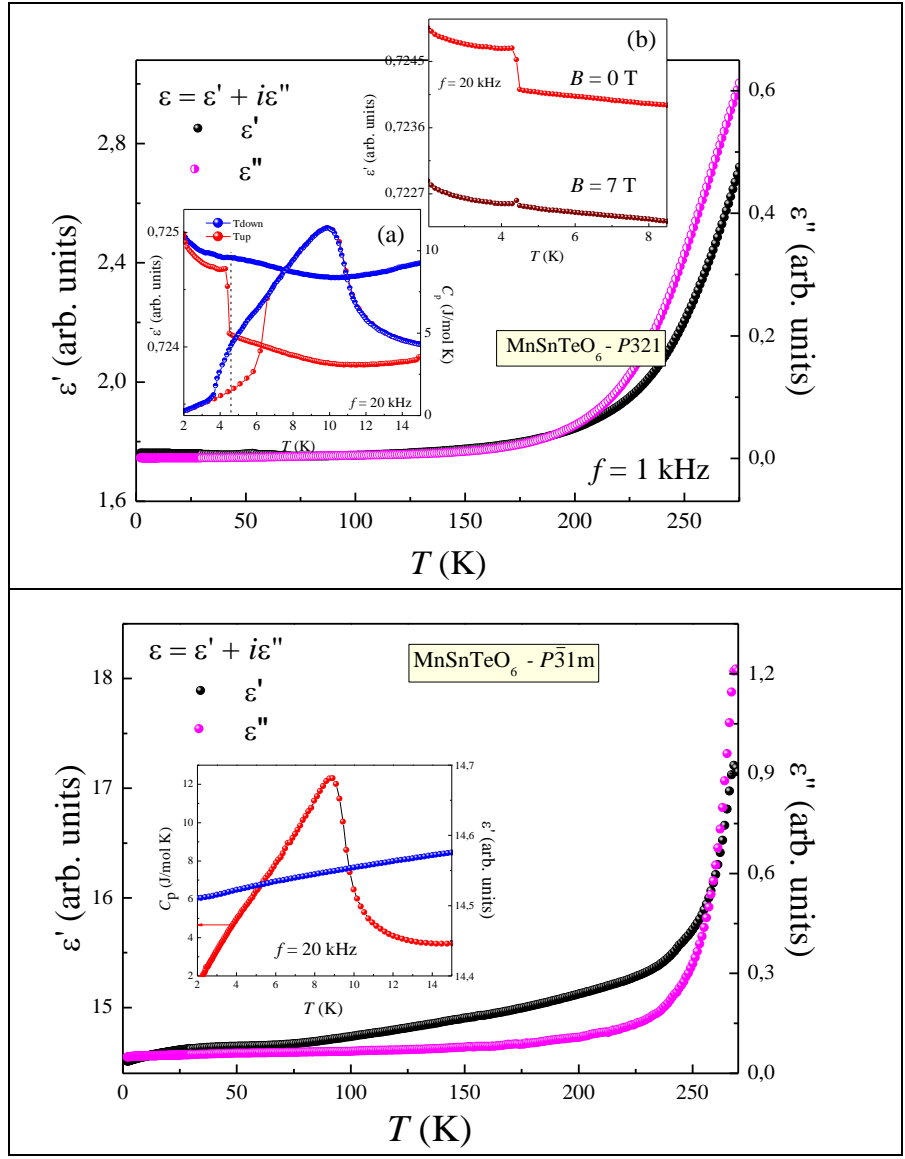

**Figure 8.** Temperature dependences of real and imaginary parts of dielectric permittivity of chiral ($P321$) (upper panel) and rosiaite-type ($P\bar{3}1m$) MnSnTeO$_6$.

### 3.6. Magnetization isotherms

Additional features may be seen on the field dependence of the magnetization, which is presented in Fig. 9. In relatively weak applied field the magnetization curves demonstrate a clear upward curvature suggesting the presence of a magnetic field induced spin-reorientation (spin-flop type) transition. The similar spin-flop-type transition in magnetic fields has also been observed for rosiaite-type ($P\bar{3}1m$) MnSnTeO$_6$.[35] These observations are characteristic of easy-axis antiferromagnets and imply a moderate magnetic anisotropy of the same scale for both polymorphs.

The critical field of the transition for chiral ($P321$) MnSnTeO$_6$ has been estimated from the maximum on the derivative curve $dM/dB(B)$ and amounts to $B_{SF} = 0.87$ T at 2 K (see inset in Fig. 9a), that is a bit less than the value determined for rosiaite-type ($P\bar{3}1m$) one (1.16 T)[35]. The corresponding jump in the magnetization data is $\Delta M_{SF} \approx 0.06\mu_B$/Mn. The position of this anomaly remains almost $T$-independent upon heating and eventually it vanishes at $T > T_N$.

It is established that the magnetization of chiral ($P321$) MnSnTeO$_6$ reaches a saturation in the moderate field: $B_{sat} = 26.1 \pm 0.2$ T at $T = 2.5$ K, and the saturation moment amounts to $4.9 \pm 0.2$ $\mu_B$/Mn which is close to the theoretically expected value for the high-spin manganese ion of 5 $\mu_B$/Mn (Fig. 9b). The values of the critical fields were used to construct the magnetic phase diagram shown in Fig. 12.

## 3.7. ESR spectroscopy

In order to probe the local magnetic properties and the spin dynamics of MnSnTeO$_6$, the ESR technique has been applied. Throughout the paramagnetic phase, the ESR absorption spectra exhibit an exchange-narrowed Lorentzian-type line corresponding to the signal from Mn$^{2+}$ ions (Fig. 10). Analysis of the ESR lineshape was carried out, taking into account that the absorption line is relatively wide. In this case, the approximating formula

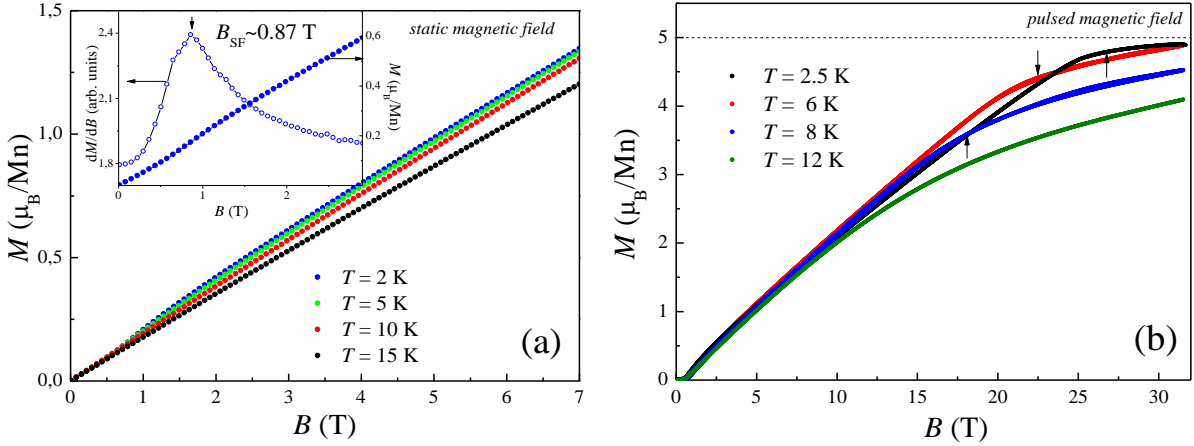

**Figure** 9. The magnetization isotherms of chiral (*P*321) MnSnTeO$_6$ measured in static (a) and pulsed magnetic fields (b). On the inset, the first derivative of the magnetization highlighting the position of the spin-reorientation transition.

includes two circulating components of a linearly polarized high-frequency field:[50]

$$\frac{dP}{dB} \propto \frac{d}{dB}\left[\frac{\Delta B}{\Delta B^2+(B-B_r)^2}+\frac{\Delta B}{\Delta B^2+(B+B_r)^2}\right] \quad (1)$$

This is a symmetric Lorentzian profile, where *P* is the power absorbed in the ESR experiment, *B* is the magnetic field, $B_r$ is the resonant field, and $\Delta B$ is the line width. The result of the fitting is shown by the red solid curves in Fig. 10. Obviously, the approximation lines agree reasonably well with the experimental data over the entire temperature range. At helium temperatures, the ESR signal degrades, indicating the establishment of a long-range antiferromagnetic order and the opening of a gap in the spectrum of magnetic excitations.

The main ESR parameters yielded from the approximation are collected in Fig. 11. The effective g-factor with typical for S-type ion Mn$^{2+}$ average value g = 1.99 ± 0.01 is practically temperature independent and demonstrates a small shift only in the proximity to $T_N$. At the same time, the linewidth increases monotonously upon lowering the temperature over the whole *T*-range. The most pronounced broadening is observed below $T \approx 150$ K indicating very wide range of the short-range correlations. The integral ESR intensity $\chi_{esr}$, which is proportional to the number of magnetic spins, was estimated by double integration of the first derivative of the experimental absorption line. The data corroborate well the behavior of the static magnetic susceptibility. The approximation of $\chi_{esr}(T)$ in accordance with the Curie-Weiss law leads to $\Theta_{esr} \approx -50$ K, which is in agreement with the negative value of the Weiss temperature estimated from $\chi(T)$ and confirms a predominance of antiferromagnetic correlations in the system.

The temperature dependence of the ESR linewidth was analyzed in the framework of two possible scenarios for the triangular antiferromagnet MnSnTeO$_6$. Firstly, the behavior of the ESR linewidth $\Delta B$ can be described in terms of the critical broadening in antiferromagnets due to slowing down of the spin fluctuations as the order-disorder temperature $T^{ESR}_N$ is approached from above. In the frame of canonical Kawasaki-Mori-Huber approach (KMH),[51-54] the temperature dependence of the $\Delta B$ can be described as follows:

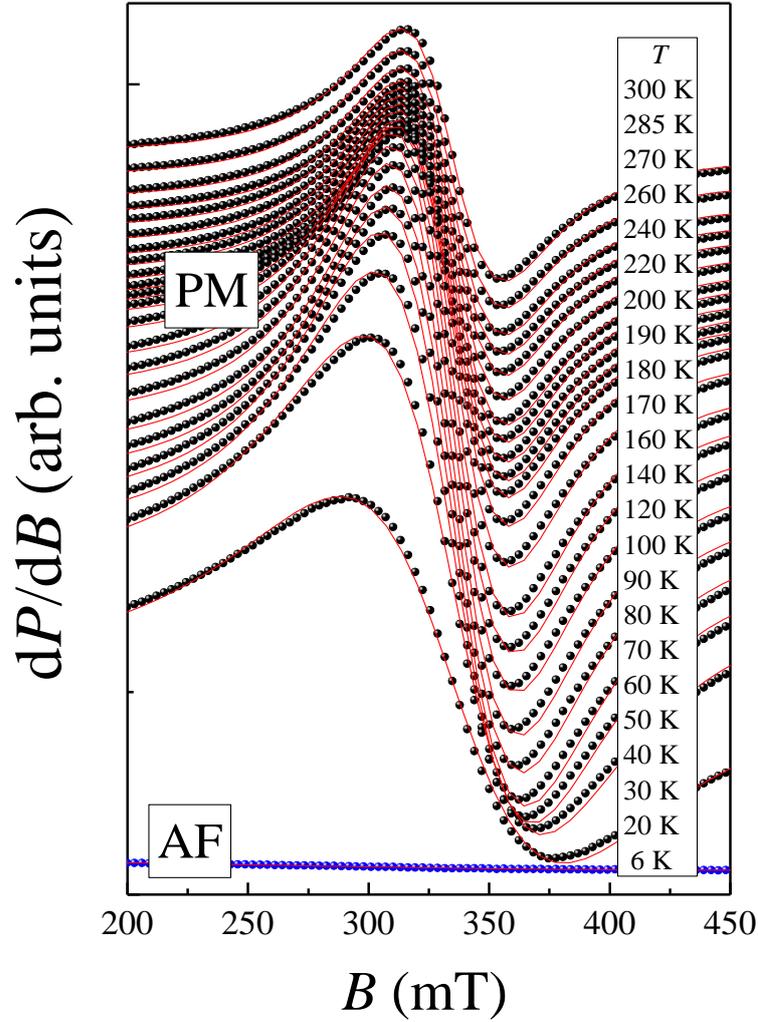

**Figure 10**. Temperature evolution of the ESR spectra of MnSnTeO$_6$: symbols are experimental data, solid curves are the result of approximation by Lorentzian profile.

$$\Delta B(T) = \Delta B^* + A \cdot \left[\frac{T_N^{ESR}}{T - T_N^{ESR}}\right]^{\beta} \quad (2)$$

where the first term $\Delta B^*$ describes the high-temperature limit of exchange-narrowed linewidth, which is temperature-independent, while the second term is responsible for the critical behavior in the vicinity of the ordering temperature $T_N^{ESR}$ and $\beta$ is the critical exponent.

Secondly, bearing in mind that according crystallographic data MnSnTeO$_6$ is a quasi-2D triangular antiferromagnet with intrinsic structural chirality, one can also apply Berezinsky-Kosterlitz-Thouless (BKT) formalism for an alternative analysis of the $\Delta B(T)$. Theoretical background predicting the formation of topological defects, Z$_2$-vortices, on 2D triangular magnetic lattice has been developed by Kawamura and Miyashita (KM).[55,56] According to KM model (modified BKT for triangular lattice case), the temperature dependence of the correlation length between the Z$_2$ vortices follows the exponential law. Being proportional to the third power of the correlation length the ESR linewidth is expressed as:

$$\Delta B = \Delta B_\infty \exp\left(\frac{3b}{\tau^\nu}\right) \quad (3)$$

where $\Delta B_\infty$ is the asymptotic high-temperature value of the linewidth, $b = \pi/2$ and $\nu = 0.37$[57] in the frame of KM approach for the triangular lattice, $\tau$ denotes normalized temperature:

$$\tau = \frac{T}{T_{KM}} - 1$$

with $T_{KM}$ being topological phase transition temperature above which the $Z_2$-vortices dissociate.

As one can see from the right panel of Fig. 11, the experimental data are better described in the framework of the first model over a wide range of the temperatures above the Néel temperature (20-300 K). Hence, the formation of vortex phases in MnSnTeO$_6$ can hardly be expected. This is in reasonable agreement with easy-axis anisotropy assumed above, which does not favor the BKT behavior. The best fit of the $\Delta B(T)$ according to Eq. (2) (red curve on the right panel of Fig. 11) resulted in the values $\Delta B^* = 33 \pm 1$ mT, $T^{ESR}_N = 6 \pm 1$ K, $\beta = 1.05$. The value $\beta$ is comparable with the values of 0.69-0.85 reported for 2D triangular multiferroics ACrO$_2$ (A=Li, H, Na)[58] but obviously higher than value of 0.39 reported for isostructural chiral MnSb$_2$O$_6$[20] that implies rather low-dimensional (2D) character of the spin dynamics in MnSnTeO$_6$.

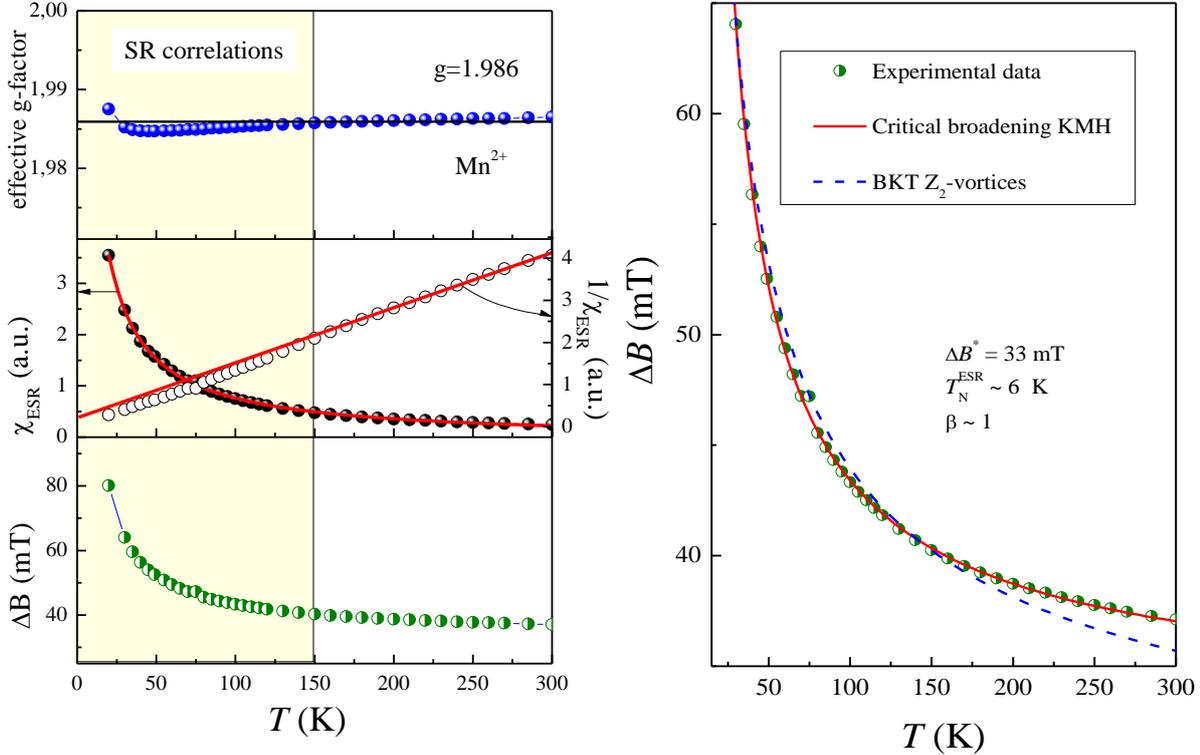

**Figure 11**. (left) Temperature dependences of ESR parameters of chiral (P321) MnSnTeO$_6$ spectra: effective g-factor (upper panel), integrated intensity (middle panel), and absorption line width (bottom panel). (right) Results of approximation of the temperature dependence of the ESR linewidth: symbols are experimental data, lines are approximations within the KMH (red solid) and KM (blue dotted) theories correspondingly.

### 3.8. Magnetic phase diagram

Summarizing the data of thermodynamic studies, a magnetic phase diagram of the new chiral magnet MnSnTeO$_6$ might be constructed (Fig. 12). At temperatures above $T_N$ in zero magnetic field the paramagnetic phase is realized. As the magnetic field increases, the phase boundary corresponding to this transition slowly shifts to lower temperature side. In addition, there is a boundary in the antiferromagnetic phase, which is probably associated with another phase transition at $T^* \approx 5$ K, and its position remains practically unaffected by a magnetic field. In addition, in external magnetic fields, about ~ 0.87 T at 2 K, a spin-reorientation transition occurs to another antiferromagnetic phase. The saturation field of the magnetization is 26 T at 2.4 K. These observations are well accounted for by both NPD studies and density functional calculations, which show that the ground state is non-collinear antiferromagnetic. According to the results of theoretical calculations and neutron data the system is ordered into an incommensurate chiral magnetic structure.

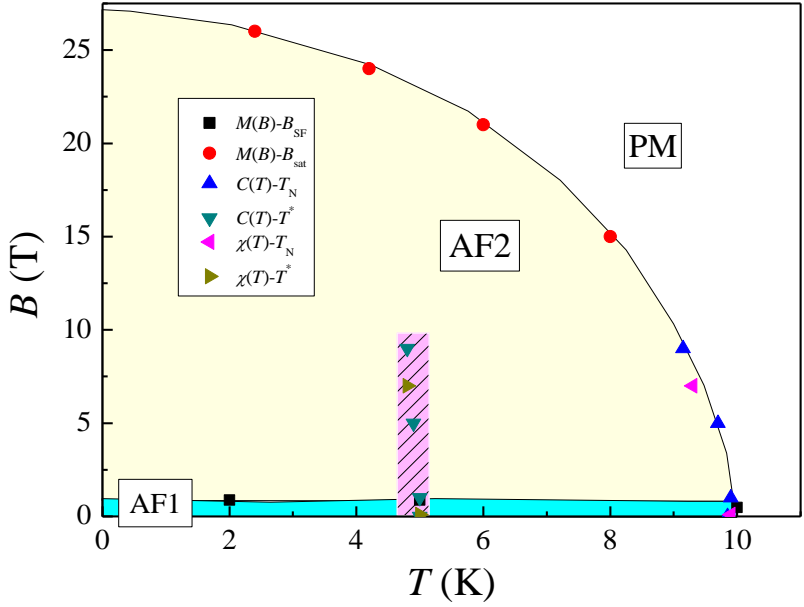

**Figure 12.** Magnetic phase diagram of chiral ($P321$) MnSnTeO$_6$.

### 3.9. Magnetic structure

The neutron diffraction patterns measured at low temperatures are shown in Fig. 13. A comparison of neutron diffraction data registered at 15 K and at lower temperatures showed the appearance of the magnetic reflections below $T_N$. Additional peaks appear in the region of the small diffraction angles, smaller than the Bragg angles of the first nuclear reflections (001) and (110), which indicates the antiferromagnetic nature of the ordering of the magnetic moments of manganese ions in the system under study. The temperature of the onset of magnetic ordering, determined from neutron data, is in good agreement with $T_N = 9.9$ K obtained from specific heat and magnetic susceptibility. The full-profile processing by Rietveld method of the experimental neutron diffraction pattern of the MnSnTeO$_6$ measured at the lowest reached temperature of 3.0 K is presented in Fig. 14.

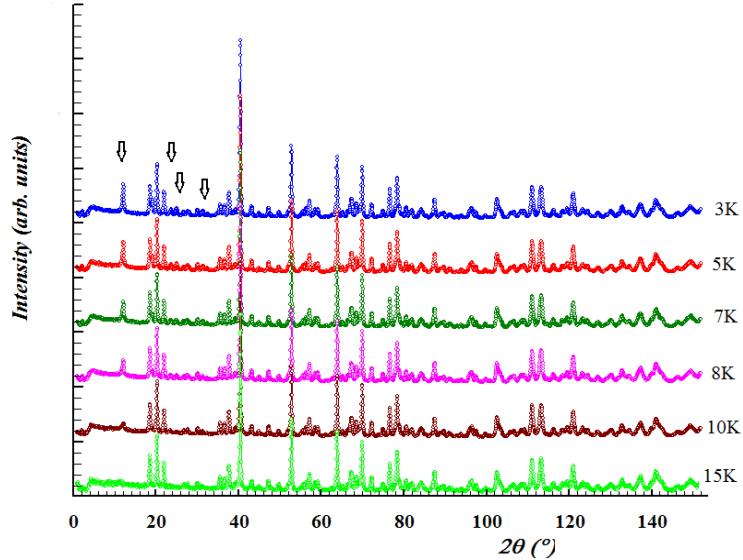

**Figure 13.** A set of low-temperature neutron diffraction patterns. Arrows are pointing the most intense additional reflections associated with magnetic scattering, antiferromagnetic kind, appearing below 15 K. Patterns are evenly shifted from each other on the y-axis to improve their perception.

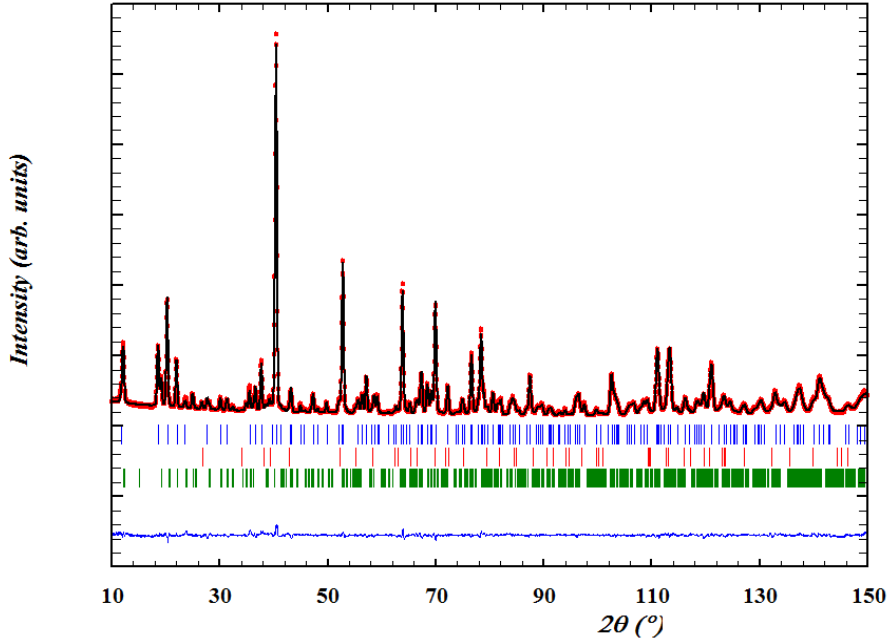

**Figure 14**. Rietveld refinement of neutron diffraction data measured at $T = 3$ K. Red dots represent experimental data, black line is calculated intensity, blue line is different curve between the observed and calculated neutron diffraction patterns, shifted down for the best perception, ticks are the position of nuclear Bragg reflections for $MnSnTeO_6$ (first blue row), impurity $SnO_2$ (second red row) and magnetic reflections of $MnSnTeO_6$ (third green row).

The positions of all additional magnetic reflections can be indexed with the propagation vector $k = (0; 0; 0.183(1))$ characterizing a spiral spin order and incommensurability of the magnetic unit cell with respect to the crystallographic one. Formation of such a structure is due to competition between different exchange interactions, which will be discussed in details in Sec. 3.10.

The model of the magnetic structure is constructed and shown in Fig. 15 as a result of the symmetry analysis and the subsequent Rietveld refinement of the neutron powder diffraction data using several models of possible spin orderings in manganese sites obtained from symmetry analysis. We can conclude that the magnetism of the compound under study is determined by triangular lattices of $Mn^{2+}$ ions ($S = 5/2$) packed along the [001] axis with a spiral spin ordering according to the obtained propagation vector $k = (0; 0; 0.183)$. So the magnetic structure in the ordered state is incommensurate. The value of the magnetic moment turned out to be $5.1(3)\mu_B/Mn$. This value practically coincides with the theoretically expected saturation magnetic moment for high-spin state of $Mn^{2+}$ ($S = 5/2$) using the value of g factor equal to 1.99 received from our experimental ESR data.

Analogue of the $MnSnTeO_6$, the trigonal $MnSb_2O_6$ has the same chiral crystallographic $P321$ space group. The ordered magnetic structure of $MnSb_2O_6$ was initially[17] determined by neutron powder diffraction as incommensurate with the crystallographic lattice and was described using the propagation vector $k = (0.013, 0.013, 0.179)$. Later, the magnetic structure of $MnSb_2O_6$ was refined and detailed in powder and single crystal experiments with significantly better resolution. The first sufficiently complete solution of the magnetic structure is given in Ref. 18, where it was described with propagation vector $k = (0; 0; 0.1820)$. This results were then confirmed[19] with the clarification that the spin-helical spiral plane in the ground state has a significant inclination from the (110) plane and that the sign of the spin-helical tilt angle is related to the spin rotation method clockwise or counterclockwise arrows, and, accordingly, with the sign of magnetically induced electric polarization. The decisive role of the anisotropy of the distorted $MnO_6$ and $SbO_6$ octahedra for stabilization of the cycloid spin structure in $MnSb_2O_6$ was noted.[18,19]

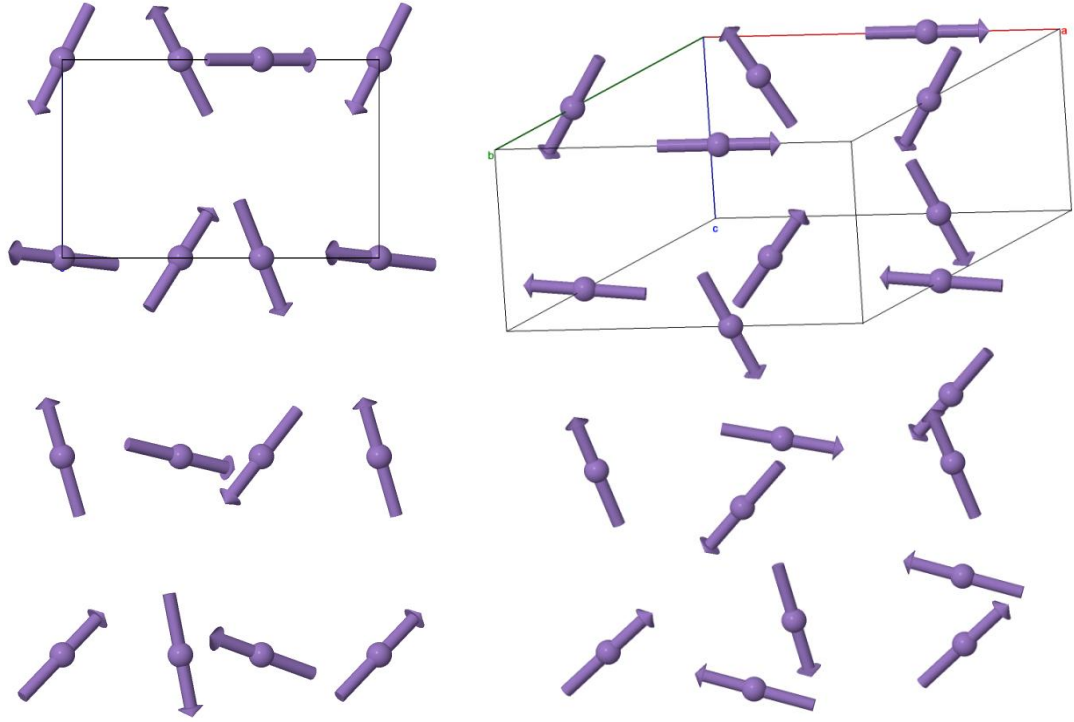

**Figure 15**. Model of the magnetic structure. The 2D distribution of the Mn spins in the *ac* plane is shown on the left, and the 3D spatial spins distribution is shown on the right. The crystallographic unit cell is also shown.

Because of the similarity in crystal structure of our new $MnSnTeO_6$ phase and its isoelectronic prototype $MnSb_2O_6$, it was natural to expect also a similar spin ordering. Therefore, a complete coincidence of the value of z-component of the propagation vectors (up to the third significant digit) in our measurements on $MnSnTeO_6$ and those performed on $MnSb_2O_6$[18,19] were surprising. The exact coincidence of the propagation vectors for spin systems $MnSnTeO_6$ and $MnSb_2O_6$ in the ground state determined from the neutron experiments indicates, among other things, that our quaternary system is absolutely ordered and there are no deviations in the distribution of atoms by their positions in comparison with the ternary $MnSb_2O_6$ system. In other words, the system investigated by us is very rigid in the sense of the magnetic structure and there is no influence of the environment of other atoms on the magnetic subsystem of manganese.

The equality of the first two components of the propagation vector to zero in comparison with small but non-zero values in the first neutron study[17] is in fact not a minor detail. It turns out that unlike the results of Ref. 17, the propagation vector ***k*** lies along [001] of the triple symmetry line, preserving the triple rotation as in langasite,[6] herewith double rotation axes lie along the directions <110> and form the compatible symmetry elements with a triple axis.

**3.10. Density functional calculations: electronic structure and exchange interaction parameters**

In order to study electronic and magnetic properties of $MnSnTeO_6$ (*P*321), we performed *ab initio* density functional theory (DFT) calculations within the spin-polarized generalized gradient approximation (GGA).[59] We used the Vienna *ab initio* simulations package (VASP).[60,61] A plane-wave cutoff energy was chosen to be 400 eV. The number of k-points depended on the supercell and varied from 20 to 200. Strong Coulomb correlations were taken into account via GGA+U method.[62] The on-site Hubbard repulsion and intra-atomic Hund's exchange parameters were chosen to be $U = 4$ eV and $J_H = 1.0$ eV (similar to typical values used for Mn in the literature [18,63]).

It was found that the band gap for $MnSnTeO_6$ is 0.8 eV. The total and partial density of states are shown in Fig. 16. The top part of the valence band is formed by strongly hybridized O 2p and Mn 3d states, while the bottom of the conduction band not by the Mn 3d, but by a mixture of Te 5s,5p and O 2p orbitals. Magnetic moments on the Mn ions were found to be 4.6 $\mu_B$, close to pure ionic value for $Mn^{2+}$.

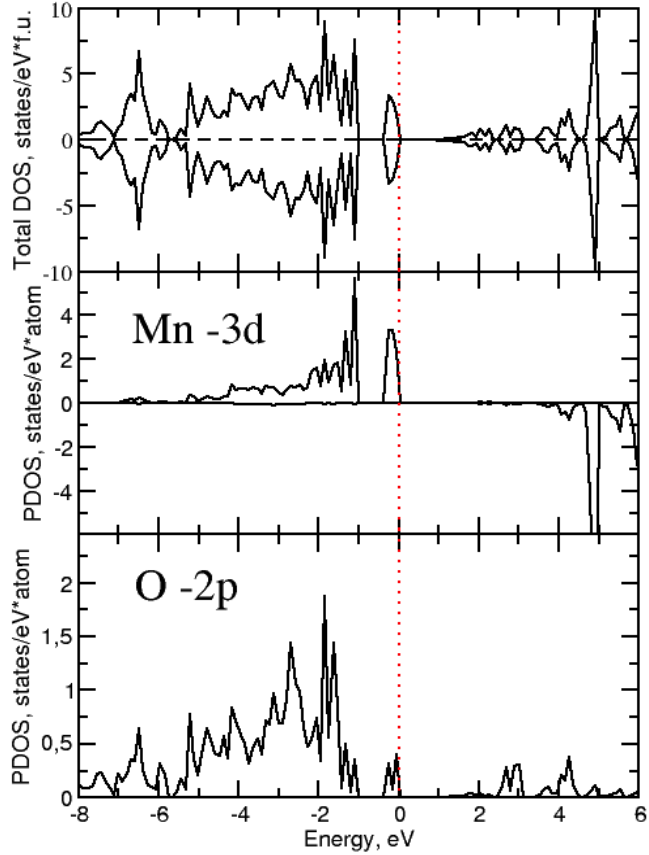

**Figure 16**. Total and partial (PDOS) density of states for $MnSnTeO_6$, as obtained in the GGA+U calculations. Positive (negative) values correspond to spin up (down). Fermi level is set to zero.

In order to study magnetic transitions, we have estimated exchange constants. They were calculated using total energies of four different collinear magnetic configurations.[18,62] A pair of interacting spins in a large supercellwas taken and changing spin projections we calculated total energies of four different states, which were used for the mapping onto the classical Heisenberg model written as following:

$$H = \sum_{i,j} J_{ij} \mathbf{S_i S_j}, \quad (4)$$

where $i, j$ – sites of the lattice, $J_{ij}$ – isotropic exchange interaction parameter between $i$ and $j$ sites, $\mathbf{S}_i$ – spin vector on the $i$-th site. We calculated seven exchange parameters. $J_1$ and $J_2$ are exchange integrals in the *ab* plane along the edges of large and small Mn triangles, respectively. $J_4$ describes coupling between different Mn planes along the *c* axis.Our calculations show that one also needs to take into account diagonal exchanges $J_3$, $J_5$, $J_6$, and $J_7$ between different layers as shown in Fig. 17. Results are summarized in Table 5.

First of all, we found that exchange constants in $MnSnTeO_6$ are very similar to those of $MnSb_2O_6$.[18] Therefore, one may expect that magnetic ground state in $MnSnTeO_6$ can be described by a set of several cycloids. The largest is the exchange interaction not in the *ab* plane or along the *c*-axis, but diagonal exchanges. Similar to $MnSb_2O_6$, the crystal structure of $MnSnTeO_6$ is chiral and there must be two different types of diagonal exchanges. The "left-handed"coupling, $J_5$ and $J_7$, are weaker than the right-handed ones, $J_3$ and $J_6$. There are two oxygen ions along $J_3$ and $J_6$ exchange paths, while there is nothing in between the Mn ions forming $J_5$ and $J_7$ paths, see Fig. 18. These two oxygen ions act as an effective media, which provide their 2p orbital for the exchange interaction. Thus, one may classify the exchange mechanism for $J_5$ and $J_7$ as super-super exchange mechanism.[64-66]

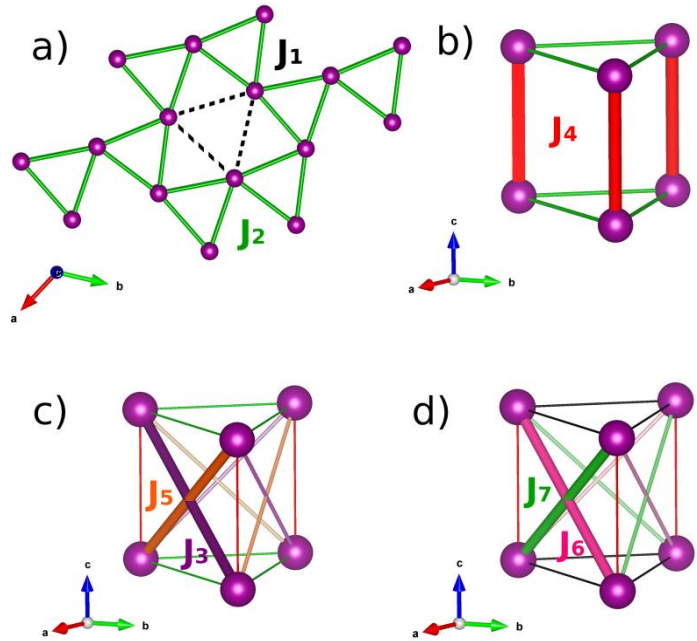

**Figure 17**. Exchange paths $J_1$–$J_7$ in MnSnTeO$_6$. In (a) large Mn$_3$ triangle are shown in black (dashed), while small ones in green (solid).

**Table 5**. Calculated values of isotropic exchange interaction parameters (in meV)

| $J_1$ | $J_2$ | $J_3$ | $J_4$ | $J_5$ | $J_6$ | $J_7$ |
|---|---|---|---|---|---|---|
| 0.12 | 0.15 | 0.60 | 0.12 | 0.02 | 0.60 | 0.02 |

We performed Monte Carlo simulations of the classical Heisenberg Hamiltonian with spin model described above using the ALPS code.[67,68] We assumed $J_3 = J_6$ and $J_1 = J_2$ and rounded them to 7 K and 2 K, respectively, and $J_4$ to 2 K. The others paths ($J_5 = J_7$) were not taken into account as they are significantly smaller. As one may see from Fig. 19, calculated within such a model magnetic susceptibility is very close to experimental one. This not only proves the adequateness of chosen spin model, but also correctness of the exchange constants calculated within the GGA+U approximation.

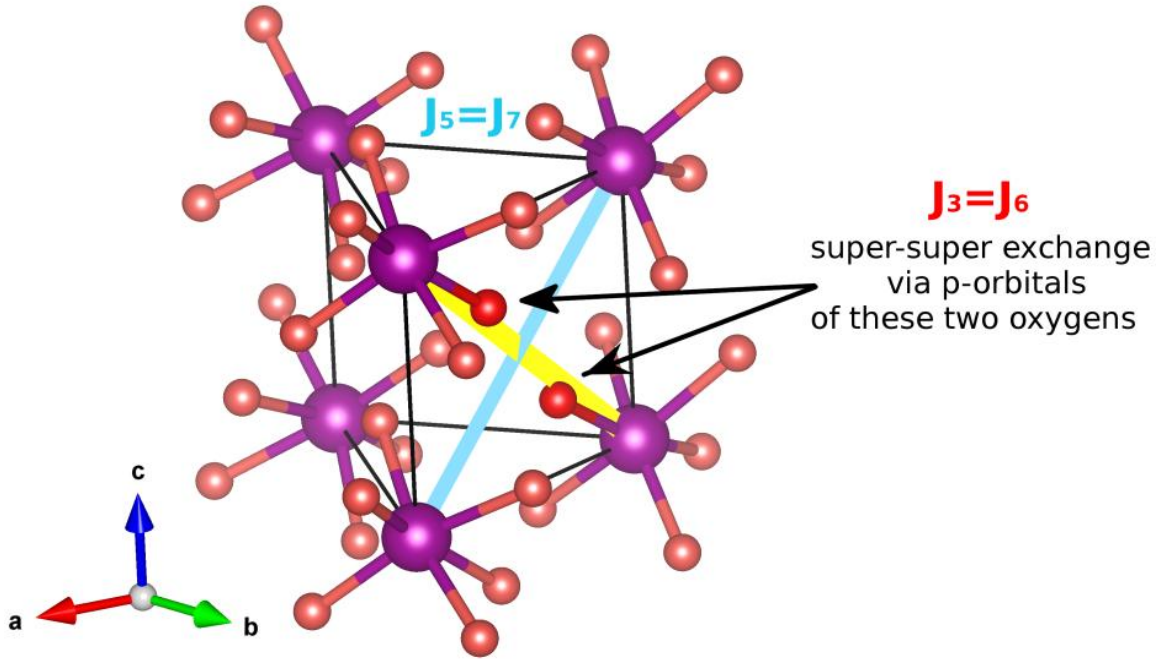

**Figure 18**. Triangular prisms, formed by the Mn ions (purple). Red balls are oxygen ions.

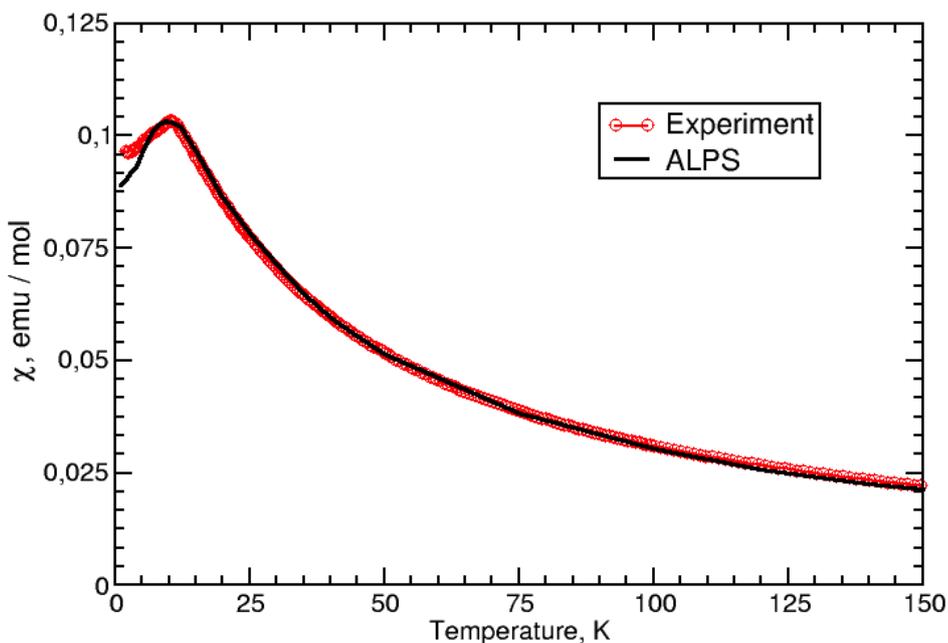

**Figure 19**. Comparison between experimental and calculated magnetic susceptibilities. Classical Heisenberg model was calculated for the spin model illustrated in Fig. 18 with exchange constants given in Table 5.

## 4. CONCLUDING REMARKS

A new triangular lattice low-dimensional magnet $MnSnTeO_6$ with chiral crystal structure (space group $P321$) was synthesized by two-stage topotactic transformation and also by direct solid-state synthesis. All thermodynamic, dielectric and resonance properties of the new polymorph are drastically different from those of the rosiaite-type ($P\bar{3}1m$) $MnSnTeO_6$.[35] In particular, the magnetic susceptibility and specific heat measurements reveal two anomalies at low temperatures. The first one at approximately 10 K is related to formation of long-range magnetic order, which was shown (by means of neutron diffraction) to have propagation vector $\boldsymbol{k} = (0,0,0.183)$. This is very similar to what was observed in closely related system $MnSb_2O_6$ earlier. Another anomaly at $T^* \approx 4.9$ K is characterized by a substantial thermal hysteresis. Dielectric permittivity was also shown to have a peak at second temperature probably suggesting formation of spontaneous electric polarization. *Ab initio* density functional theory calculations demonstrate that, in contrast to naive expectations, the largest exchange coupling is not between Mn ions in the same triangular plane, but between planes along diagonals. All exchange parameters are antiferromagnetic and reveal moderate frustration.

## ASSOCIATED CONTENT
**Supporting Information**
The Supporting Information is available free of charge on the ACS Publications website. Details of the sample preparation; Details of the XRPD data collection and structure refinement; Atomic positions and thermal displacement parameters, Principal interatomic distances, bond valences and bond angle ranges; Refinement of Mn occupancy; Temperature dependence of the magnetic susceptibility: comparison of two $MnSnTeO_6$ polymorphs; Generalized spin-model.

## Accession codes
CCDC - 1943127 contains the supplementary crystallographic data for this paper. These data can be obtained free of charge via www.ccdc.cam.ac.uk/data_request/cif, or by emailing data_request@ccdc.cam.ac.uk, or by contacting The Cambridge Crystallographic Data Centre, 12 Union Road, Cambridge CB2 1EZ, UK; fax: +44 1223 336033**.**

**ACKNOWLEDGMENTS**
The reported study was funded by Russian Science Foundation according to the research projects № 18-12-00375 (A.I.K. and M.D.K.) for neutron studies and 17-12-01207 (E.A.Z. and S.V.S.) for magnetic, dielectric and specific heat studies as well as theoretical calculations. Sample preparation and diffraction studies by M.A.E., M.D.K., A.I.K., and V.B.N. were supported by the grant 18-03-00714 from the Russian Foundation for Basic Research. V.B.N. thanks the International Centre for Diffraction Data for Grant-in-Aid 00-15. A.N.V. and S.V.S. acknowledge the support by the Russian Ministry of Education and Science of the Russian Federation through NUST «MISiS» grant K2-2017-084 and by the Act 211 of the Government of Russia, contracts 02.A03.21.0004, 02.A03.21.0011 and 02.A03.21.0006. The authors thank Dr. Yu.V. Popov (SFU's Shared Use Centre "Research in Mineral Resources and Environment") for the EDX analysis.

Supporting Information for the paper
# MnSnTeO$_6$: a Chiral Antiferromagnet Prepared by a Two-Step Topotactic Transformation

**Contents**

## S1. Preparation of samples
### S1.1. Starting materials

Only reagent-grade chemicals were used. $MnSO_4 \cdot xH_2O$, $SnO_2$ and $TeO_2$ were calcined in air for 2 h at 400 °C, and $Mn_2O_3$, at 750 °C. Hydrous $MnCl_2$ was dried in vacuum at 200 °C. $Na_2CO_3$, KCl, $Li_2SO_4$ and CsCl were dried in air at 150 °C. All dried materials were stored in closed containers in a desiccator. Metallic tin and amorphous $TeO_3$ were used as received.

$Na_2SnTeO_6$ was prepared by solid-state reactions as reported earlier[32] and its phase purity was verified by PXRD.

### S1.2. Ion exchange

In absence of detailed phase diagrams for the multicomponent reciprocal salt systems, various salt mixtures based on $MnSO_4$ were tested to reduce its melting point.[32] The Sample 1 used for structural and physical studies in this work was prepared from the following mixture:

$Na_2SnTeO_6 + 5\ MnSO_4 + Li_2SO_4 + 7\ KCl + 5CsCl$

Five-fold excess of $MnSO_4$ was intended to provide complete Mn substitution for sodium; to the same end, we avoided sodium salts as flux components. Powdered $Na_2SnTeO_6$ and salts were weighed and mixed rapidly (to avoid hydration in air) and placed in a glass test tube inserted into a vertical tubular furnace. The upper cold end of the test tube was closed with a stopper having thin inlet and outlet tubes for gases. After passing considerable excess of argon to replace air, the furnace was turned on, heated to 430 °C and held for an hour at this temperature under reduced flow of argon. Then, the tube was extracted from the furnace and cooled, still under flowing argon. The cooled mixture was washed with warm distilled water several times until negative reaction to sulfate ion. Then the precipitate was dried at 150 °C. As indicated in the main text, the sample was mostly rosiaite-type with admixtures of $SnO_2$ and targeted P321 polymorph, but after an hour at 700 °C the rosiate-to-P321 transition was completed.

### S1.3. Direct preparation of $MnSnTeO_6$ by solid-state synthesis

For the co-precipitation route, stoichiometric amounts of $Mn_2O_3$ and metallic tin were dissolved jointly in strong hydrochloric acid. Thus obtained solution was added dropwise, under vigorous stirring, to an excess of cooled aqueous solution of ammonia with hydrogen peroxide added to convert Sn(2+) to Sn(4+). Due to fast decomposition of the peroxide under catalytic action of manganese hydroxide, it was necessary to add, from time to time, new portions of $H_2O_2$. The precipitate was vacuum-filtered, washed with water until negative reaction to chloride ion, and dried at room temperature. Then, it was mixed with the stoichiometric amount of amorphous $TeO_3$, pressed, calcined for 2 h at 300 °C, reground, pressed again, covered with packing powder of the same composition, held for 4 h at 750 °C in a covered crucible and cooled.

## S2. Structural data

Table S1. Details of the XRPD data collection and structure refinement of $MnSnTeO_6$

| Crystal system | | Trigonal | Density (calc.) | | 6.179 |
|---|---|---|---|---|---|
| Space group | | P 3 2 1 (no. 150) | Texture parameters (March-Dollase) | | axis 001 ratio 1.00 |
| Lattice constants | $a$, Å | 8.78366(10) | 2Θ range, ° | | 15.02–106.06 |
| | $b$, Å | 8.78366(10) | 2Θ step, ° | | 0.02 |
| | $c$, Å | 4.79333(9) | No. of data points | | 4553 |
| | $\gamma$, ° | 120.0 | No. of reflections calc. ($\alpha_1$ only) | | 372 |
| Cell volume, Å$^3$ | | 320.272(6) | | | |
| Formula weight | | 397.22 | No. of variables | | 52 |
| Z | | 3 | Agreement factors | GOF | 1.40 |
| Wavelengths, Å | $\alpha_1$ | 1.5406 | | $R(F^2)$ | 0.05468 |
| | $\alpha_2$ | 1.5443 | | $R_p$ | 0.0749 |
| | Ratio | 0.5 | | $R_{wp}$ | 0.1115 |
| Polarization ratio | | 0.707 | | $\chi^2$ | 1.98 |

Table S2. Atomic positions and thermal displacement parameters in MnSnTeO$_6$ from room-temperature XRPD

| Atom | Site | x/a | y/b | z/c | occupancy | U (Å$^2$) |
|---|---|---|---|---|---|---|
| Te1 | 1a | 0 | 0 | 0 | 1 | 0.01341(9) |
| Te2 | 2d | 1/3 | 2/3 | 0.50157(17) | 1 | 0.01278(7) |
| Sn | 3f | 0.29875(19) | 0 | 1/2 | 1 | 0.01262(6) |
| Mn | 3e | 0.62957(4) | 0 | 0 | 1 | 0.00913(10) |
| O1 | 6g | 0.09566(21) | 0.8825(15) | 0.78802(17) | 1 | 0.00673(5) |
| O2 | 6g | 0.47129(22) | 0.59275(16) | 0.73111(19) | 1 | 0.00701(4) |
| O3 | 6g | 0.21519(18) | 0.77075(27) | 0.2897(20) | 1 | 0.01506(5) |

Table S3. Principal interatomic distances (L), bond valences (BV) and bond angle ranges in MnSnTeO$_6$ from room-temperature XRPD

| Bond | L (Å) | BV | angles | angle range (°) |
|---|---|---|---|---|
| Te1-O | 1.9159(10)×6 | 1.016 | O-Te1-O | 81.4 – 94.5 |
| Te2-O | 1.9716(15)×3 | 0.880 | O-Te2-O | 81.4 – 95.9 |
|  | 1.973(5)×3 | 0.876 |  |  |
| Sn-O | 2.033(5)×2 | 0.729 | O-Sn-O | 77.7 – 104.7 |
|  | 2.071(2)×2 | 0.633 |  |  |
|  | 2.077(9)×2 | 0.621 |  |  |
| Mn-O | 2.128(2)×2 | 0.394 | O-Mn-O | 69.5 – 98.7 |
|  | 2.167(6)×2 | 0.359 |  |  |
|  | 2.193(9)×2 | 0.338 |  |  |

Table S4. Refinement of Mn occupancy on the 15 K NPD data

| Mn occupancy | R$_p$ | R$_{wp}$ | R$_{exp}$ | $\chi^2$ |
|---|---|---|---|---|
| Fixed: 1 | 9.53 | 9.37 | 7.68 | 1.49 |
| Refined: 0.992(16) | 9.64 | 9.27 | 7.71 | 1.51 |

**S3. Temperature dependence of the magnetic susceptibility: comparison of two MnSnTeO$_6$ polymorphs.**

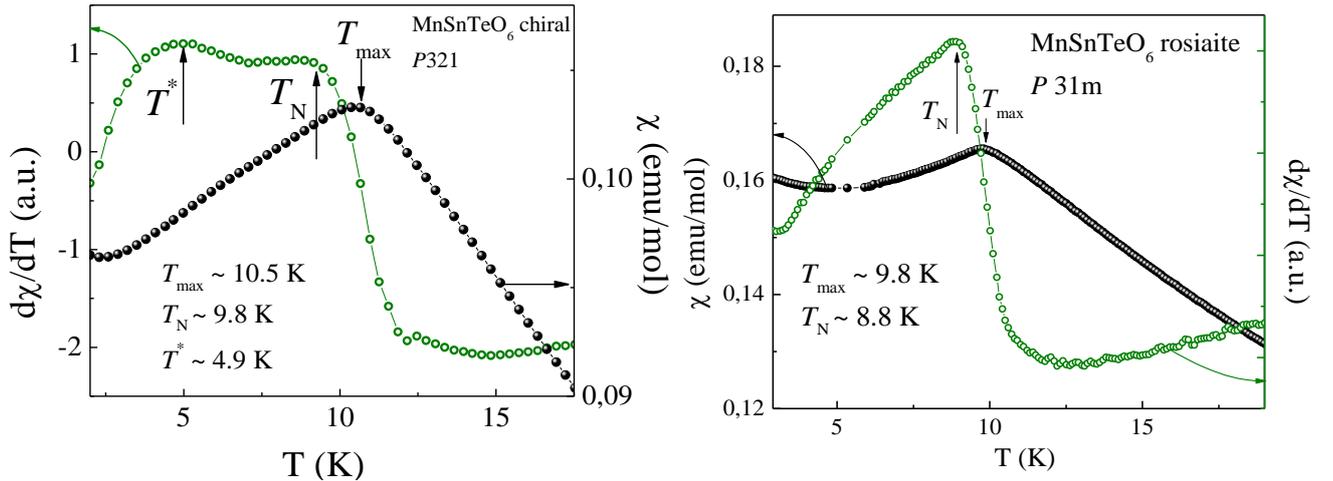

**Figure S1.** Low temperature part of the χ(*T*) in comparison with its derivative dχ/d*T* (*T*) for two polymorphs MnSnTeO$_6$: stable P321 (present work) and metastable P31m (analysis of the data from Ref. 32). Results indicate the presence of additional anomaly at *T**~ 4.9 K for stable P321 compound.

## S4. Generalized spin-model

As it was mentioned in the main part of the paper, the spin-Hamiltonian can be substantially simplified, if we assume $J_1=J_2$, $J_3=J_6$, and neglect $J_5$ and $J_7$. Then the spin model can be described as three interpenetrating simple cubic lattices, see Fig. S2(a) and (b). Each of the lattices is given by $J_6$ (= $J_3$) exchange parameter (right-handed diagonal exchanges in the Mn prisms). These cubic lattices are coupled by frustrating $J_1$ (= $J_2$) exchange (exchanges in the $ab$ plane), see Fig. S2(c), and by non-frustrating $J_4$ (along the $c$ axis), Fig. S2(d).

The spin model described above with three exchange parameters can be applied not only to $MnSnTeO_6$, but also to other materials with similar crystal structure (e.g. $MnSb_2O_6$). Therefore, it is interesting to study how magnetic properties change if one varies exchange parameters. We fixed non-frustrating $J_4 = 2$ K (like in the previous calculations), but changed the ratio between exchange parameters in the cubic lattices ($J_1$) and frustrating exchange ($J_6$): $\beta_J = J_1/J_6$. The total exchange determined by Curie-Weiss parameter $q=\sum_i N_i J_i$, where $N_i$ is a number of $i$-th exchange interaction $J_i$, was kept to be the same.

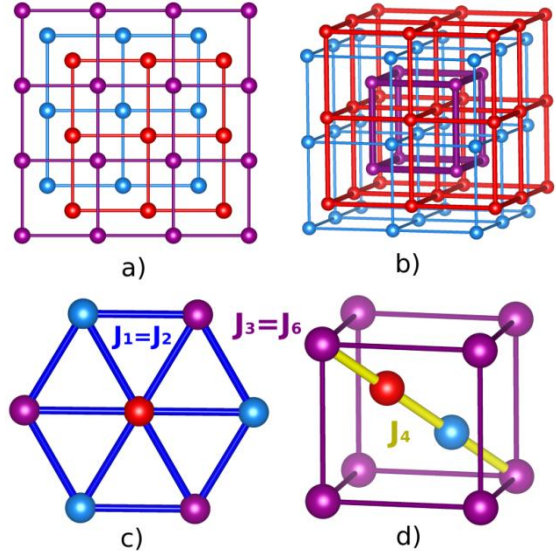

**Figure S2**. Simplified spin model, which can be used to describe magnetic properties of $MnAA'O_6$ compounds where A and A′ are p-elements (Sn, Sb or Te). It can be rationalized as three interpenetrating simple cubic lattices linked by two additional frustrating $J_1$ and non-frustrating $J_4$ exchange parameters.

Fig. S3 illustrates how temperature $T_{max}$ corresponding to the maximum of magnetic susceptibility changes with $\beta_J$. One may fit obtained $T_{max}/\beta_J$ dependence with hyperbola $T_{max}/\beta_J = c/(\beta_J+b)$ with coefficients $c = 9.730$ K and $b = -0.003$, or alternatively $T_{max} = c(1-b/\beta_J)$. Thus, we see that since coefficient $b$ is rather small the frustration (i.e. $\beta_J = J_1/J_6$) would not strongly affect $T_{max}$.

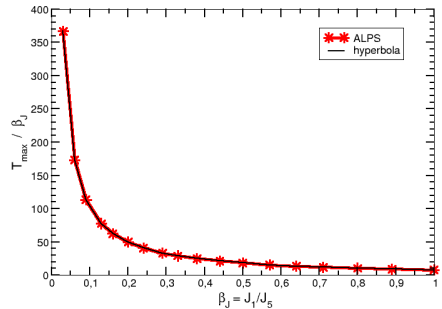

**Figure S3**. Dependence of temperature $T_{max}$ of corresponding to maximum of magnetic susceptibility on the $\beta_J = J_1/J_6$ ratio. Results of the classical Heisenberg Hamiltonian simulation of the three parameters ($J_1$, $J_6$, and $J_4$) and their hyperbolic fit.